\documentclass[lettersize,journal]{IEEEtran}

\usepackage{ifluatex}
\usepackage{ifpdf}

\ifluatex
	\usepackage{fontspec}
	\usepackage{anyfontsize}
	\usepackage{amsmath, amsfonts, amsthm}
	
	\usepackage{unicode-math}
	
\newfontfamily\tgpfont{TeX Gyre Pagella}

\newfontfamily\arialfont{Arial}

\else
\ifpdf
	\usepackage[utf8]{inputenc}
	\usepackage{fontenc}
	\usepackage{anyfontsize}
	\usepackage{amsmath, amsfonts, amsthm}
	\usepackage{upgreek}
	\NewDocumentCommand{\symbfup}{m}{%
		\boldsymbol{\mathrm{#1}}%
	}
	\NewDocumentCommand{\symbfsfup}{m}{%
		{\bfseries\sffamily #1}%
	}
	\NewDocumentCommand{\symsfup}{m}{%
		{\textsf{#1}}%
	}
	\NewDocumentCommand{\symbb}{m}{%
		\mathbb{#1}%
	}
	\NewDocumentCommand{\symup}{m}{%
		\text{#1}%
	}
\fi
\fi
\usepackage[svgnames]{xcolor}
\usepackage[hidelinks]{hyperref}
\usepackage{thmtools}
\usepackage[nameinlink]{cleveref}
\usepackage{lipsum}
\usepackage{xparse}
\usepackage[math]{cellspace}
\usepackage{nicefrac}
\usepackage{letltxmacro}
\usepackage[breakable]{tcolorbox}
\tcbuselibrary{skins,breakable}
\tcbset{shield externalize}
\usepackage[
	backend=biber,
	style=ieee,
	doi=false,
	isbn=false,
	url=false,]{biblatex}
\usepackage{relsize}
\usepackage{tikz}
\usepackage{pgfplots}
\usepackage{caption}
\usepackage{subcaption}
\usepackage{titlesec}

\usepackage{booktabs}
\usepackage{siunitx}
\usepackage{derivative}
\usepackage{longtable}

\usepackage{algorithm}
\usepackage{algpseudocode}
\usepackage{algorithmicx}
\usepackage{array}
\usepackage{textcomp}
\usepackage{stfloats}
\usepackage{pgfplotstable}
\usepackage{hhline}
\usepackage{supertabular}

\usepackage{import}
\usepackage{ipsum}
\usepackage{orcidlink}
\usepackage{tikz-cd}
\usepackage[scr=txupr]{mathalpha}
\usepackage{tocloft}

\usepackage[normalem]{ulem}

\NewDocumentCommand{\IfEmpty}{m m G{}}{\ifthenelse{\equal{#1}{}}{#2}{#3}}%
\let\ABD\AtBeginDocument
\let\nfrac\nicefrac
\NewDocumentCommand{\MakeSymbol}{m m}{
    \ABD{\DeclareDocumentCommand{#1}{}{#2}}
}
\makeatletter
\definecolor{TodoBlue}{HTML}{50b0d0}
\definecolor{TodoGreen}{HTML}{50d060}
\definecolor{TodoOrange}{HTML}{d09030}
\DeclareDocumentCommand{\@todo}{m O{TodoOrange}}{%
	\noindent{\small\texttt{\textcolor{#2}{-TODO-:} [\\
	\phantom{aaaa}\begin{minipage}[t]{0.8\linewidth}\noindent{\color{#2}#1}\end{minipage}\\
	]}}\\%
}
\DeclareDocumentCommand{\todotext}{m}{%
	\@todo{#1}[TodoBlue]
}
\DeclareDocumentCommand{\todosym}{m}{%
	\@todo{#1}[TodoGreen]
}
\let\todo\@todo
\makeatother

\ABD{%
	\def\checkmark{
    \tikz\fill[scale=0.4](0,.35) -- (.25,0) -- (1,.7) -- (.25,.15) -- cycle;} 
	\NewDocumentCommand{\done}{s m}{%
		\IfBooleanTF{#1}{
			{\color{TodoBlue!30!white} \sout{#2}}%
		}{%
			{\color{TodoGreen!30!white}\checkmark \sout{#2}}%
		}%
	}
}

\DeclareDocumentCommand{\re}{}{%
    \symbb{R}%
}
\DeclareDocumentCommand{\im}{}{%
    \symbb{I}%
}

\AtBeginDocument{%
\DeclareDocumentCommand{\Re}{s m}{%
	\IfBooleanTF{#1}{%
		{#2}^{\re}%
	}{%
		#2^{\re}%
	}%
}
\DeclareDocumentCommand{\Im}{s m}{%
	\IfBooleanTF{#1}{%
		{#2}^{\im}%
	}{%
		#2^{\im}%
	}%
}
}

\DeclareMathOperator*{\argmin}{argmin}

\DeclareDocumentCommand{\expec}{o m}{%
	\IfValueTF{#1}{\underset{#1}{\mathbb{E}}}{\mathbb{E}}\left\{#2\right\}
}

\DeclareDerivative{\npdv}{\partial}[style-var=multiple, style-var-/=multiple, style-var-!=mixed, style-var-/!=multiple, delims-eval=(), delims-eval-/=(), delims-eval-!=(),style-frac=\nfrac]

	\MakeSymbol{\j}{\text{j}}

\MakeSymbol{\minus}{
    \text{-}
}
\MakeSymbol{\dg}{^{\symsfup{o}}}
\MakeSymbol{\eqc}{\text{,}}
\MakeSymbol{\eqp}{\text{.}}
\MakeSymbol{\eqsc}{\text{;}}
\MakeSymbol{\dd}{\symsfup{d}}

\MakeSymbol{\defas}{%
    \mathbin{\oset{\raisebox{1pt}{\scalebox{0.6}[0.6]{$\triangle$}}}{=}}%
}
\MakeSymbol{\div}{%
    \mathbin{//}%
}
\MakeSymbol{\mod}{%
    \mathbin{\%}%
}
\MakeSymbol{\kp}{%
    \mathbin{\otimes}%
}
\MakeSymbol{\conv}{%
    \mathbin{\ast}%
}
\MakeSymbol{\bconv}{%
	\mathbin{\circledast}%
}
\MakeSymbol{\vbconv}{%
	\mathbin{\rharpoon{\circledast}}%
}
\MakeSymbol{\e}{%
	\mathrm{e}%
}
\MakeSymbol{\naturals}{%
	\mathbb{N}%
}

\MakeSymbol{\PM}{
	\mathbin{\ooalign{\raisebox{1.5pt}{\scalebox{0.8}{$+$}}\cr\hfil\raisebox{-1.5pt}{\scalebox{0.8}{$-$}}\hfil}}
}

\MakeSymbol{\bsquare}{\fcolorbox{black}{black}{\null}}

\DeclareDocumentCommand{\review}{m g}{
    \item \IfValueTF{#2}{
         {\color{LightGray}\soutit{#1}} - #2
    }{
        {\color{Gray}\textit{#1}}
    }
}

\newcounter{reviewer}
\DeclareDocumentEnvironment{reviews}{}{
    \begin{enumerate}[label=\thereviewer.{\alph*}]
    \footnotesize
}{
    \end{enumerate}
}
\definecolor{ColA}{HTML}{991F3D}
\definecolor{ColB}{HTML}{997A1F}
\definecolor{ColC}{HTML}{3D991F}
\definecolor{ColD}{HTML}{1F997A}
\definecolor{ColE}{HTML}{1F3D99}
\definecolor{ColF}{HTML}{7A1F99}
\crefname{equation}{Eq.}{Eqs.}
\crefname{figure}{Fig.}{Figs.}
\crefname{table}{Table}{Tables}
\crefname{algorithm}{Algorithm}{Algorithms}
\crefname{section}{Section}{Sections}
\crefname{paragraph}{Paragraph}{Paragraph}

\Crefname{figure}{Figure}{Figures}

\creflabelformat{paragraph}{(#2#1#3)}

\makeatletter
\ABD{
	\DeclareDocumentCommand{\mref}{m g}{%
		\IfValueT{#2}{\cite{#2}-}\cref{#1}
	}%
}%
\ABD{
}%
\makeatother

\DeclareDocumentCommand{\extlabel}{m m O{0}m}{%
	\newlabel{#1}{{20}{1}{}{#2.#3.#4}{}}
	\newlabel{#1@cref}{{[#2][#4][]#4}{[1][1][]1}}
}

\usepgfplotslibrary{polar}
\newlength{\Yaxisshift}
\NewDocumentCommand{\ppref}{m}{%
	\ref*{#1}%
}

\usepgfplotslibrary{colorbrewer}
\NewDocumentEnvironment{heatmap}{m O{}}{
	\begin{axis}[
		colormap/viridis,
		colorbar horizontal,
		colorbar style={
			colormap/viridis,
			width=0.6\linewidth,
			xlabel = Beampattern,
			xtick = {-10, -20, ..., -30},
			xticklabel style={
				/pgf/number format/.cd,
				fixed,
				precision=0,
				fixed zerofill,
			},
			major tick length=4pt,
			xtick style={black},
			xtick pos=top,
			extra x ticks = {0, \ymin},
			extra x tick style = {major x tick style={draw=none}},
			point meta min=\ymin,
			point meta max=0,
		},
		width=1\linewidth,
		height=0.7\linewidth,
		point meta min=\ymin,
		point meta max=0,
		mesh/cols=#1,
		xlabel style = {anchor=north, yshift=0.8\linewidth-2.0em},
		view={0}{90},
		xtick = {-180, -90, ..., 180},
		xtick style = {black},
		xticklabel style = {yshift=-3pt},
		xticklabel={$\pgfmathprintnumber{\tick}\dg$},
		grid style = {draw=none},
		ylabel style={yshift=-\linewidth+0.2em},
		ytick = {0, 8, ..., 32},
		major tick length=2pt,
		ytick style = {black},
		ytick pos = left,
		mesh/ordering=x varies,
		#2
		]
		\centering
	}{
	\end{axis}
}

\DeclareDocumentCommand{\heatmapfig}{O{0.8\linewidth} O{#7} O{\meshcols} O{\meshrows} m m m O{}}{
	\begin{subfigure}{#1}
		\centering
		\begin{tikzpicture}
			\begin{heatmap}{\meshcols}[colorbar to name={#2}, #8]
				\addplot3[surf, mesh/cols=#3, mesh/rows=#4, shader=interp] table[x=ang, y=freq, z=val, col sep=comma] {#5};
			\end{heatmap}
		\end{tikzpicture}
		\vspace*{-2mm}\caption{#6}
		\label{subfig:#7}
		\vspace*{2mm}
	\end{subfigure}
}

\newlength{\wdtwenty}
\pgfplotsset{
	every colorbar global/.append style={
		zmin=,zmax=,
	}
}

\NewDocumentEnvironment{lineplot}{O{axis} m m m}{ 
	\begin{#1}[
		width=0.9\linewidth,
		height=0.6\linewidth,
		xtick pos=bottom,
		ytick pos=left,
		xlabel = #2,
		ylabel = #3,
		yticklabel style={text width=\wdtwenty, align=right},
		ylabel style={yshift=-1em},
		legend cell align={left},
		legend style={
			fill=white,
			draw opacity=1,
			text opacity=1,
			legend columns=2,
			/tikz/every even column/.append style={column sep=1em}
		},
		#4,
		]
		\draw[thin, dash pattern = {on 4pt off 1pt}] (axis cs:\pgfkeysvalueof{/pgfplots/xmin},0) -- (axis cs:\pgfkeysvalueof{/pgfplots/xmax},0);
	}{
	\end{#1}
}
\NewDocumentEnvironment{boxplot}{O{}}{ 
	\begin{axis}[
		width=0.9\linewidth,
		height=0.45\linewidth,
		boxplot/draw direction = y,
		xtick pos=bottom,
		ytick pos=left,
		xtick style = {draw=none},
		yticklabel style={text width=\wdtwenty, align=right},
		ylabel style={yshift=-1em},
		#1,
		]
	}{
	\end{axis}
}
\NewDocumentEnvironment{angleboxplot}{O{}}{ 
	\begin{semilogyaxis}[
		scale=1.15,
		width=0.85\linewidth,
		height=110pt,
		boxplot/draw direction = y,
		xtick pos=bottom,
		ytick pos=left,
		xtick style = {draw=none},
		yticklabel style={text width=\wdtwenty, align=right},
		ylabel style={yshift=-1em},
		ylabel={Angle error ($\dg$)},
		xticklabel style = {align=center, font=\small},
		ymin=0.1,
		ymax=165,
		ytick={0.1, 1, 10, 100},
		yticklabels={0.1, 1, 10, 100},
		ymajorgrids,
		grid style={line width=.1pt, draw=gray!10},
		major grid style={line width=.2pt,draw=gray!50},
		#1,
		]
	}{
	\end{semilogyaxis}
}

\NewDocumentCommand{\addboxplot}{m m m m m O{} O{}}{%
	\addplot+[forget plot, solid, thick, boxplot prepared={%
		lower whisker=#1,%
		lower quartile=#2,%
		median=#3,%
		upper quartile=#4,%
		upper whisker=#5,%
		#6,
	},%
	#7,fill=white,fill opacity=0.6] coordinates{};%
}

\NewDocumentEnvironment{timeplot}{m O{}}{
	\begin{lineplot}{Time ($\si{\second}$)}{#1}{0}{7.2}{#2}
	}{
	\end{lineplot}
}

\NewDocumentEnvironment{snrplot}{m O{}}{
	\begin{lineplot}{SER ($\dB$)}{#1}{-30}{10}{#2}
	}{
	\end{lineplot}
}

\NewDocumentEnvironment{errplot}{m O{}}{
	\begin{lineplot}[semilogxaxis]{Error}{#1}{0}{10}{log basis x=10,#2}
	}{
	\end{lineplot}
}

\NewDocumentEnvironment{lyplot}{m O{}}{
	\begin{lineplot}{X}{#1}{1}{16}{xtick={1, 4, ..., 16}, #2}
	}{
	\end{lineplot}
}

\NewDocumentEnvironment{nbinsplot}{m O{}}{
	\begin{lineplot}{Number of bins}{#1}{32}{512}{xtick={32, 128, ..., 512}, #2}
	}{
	\end{lineplot}
}

\tikzstyle{axis} = [
black
]
\tikzstyle{styleA} = [
	ColA!90!black, 
	dash pattern = {on 5pt off 0pt},
	very thick
]
\tikzstyle{styleB} = [
	ColB!90!black, 
	dash pattern = {on 1.2pt off 0.6pt},
	very thick
]
\tikzstyle{styleC} = [
	ColC!90!black, 
	dash pattern = {on 4pt off 1.2pt},
	very thick
]
\tikzstyle{styleD} = [
	thick, 
	ColD!90!black
]
\tikzstyle{styleE} = [
	ColE!90!black, 
	dash pattern = {on 4pt off 1.2pt on 1.2pt off 1.2pt},
	very thick
]
\tikzstyle{styleF} = [
	thick, 
	ColF!90!black
]

\tikzstyle{resA} = [
thick, ColA, dash pattern = {on 1pt off 0pt},
]
\tikzstyle{resB} = [
thick, ColB, dash pattern = {on 2pt off 1pt},
]
\tikzstyle{resC} = [
thick, ColC, dash pattern = {on 4pt off 2pt},
]
\tikzstyle{resD} = [
thick, ColD, dash pattern = {on 2pt off 3pt},
]
\tikzstyle{resE} = [
thick, ColE, dash pattern = {on 2pt off 1pt},
]
\tikzstyle{resF} = [
thick, ColF, dash pattern = {on 3pt off 2pt},
]
\tikzstyle{resG} = [
thick, ColG, dash pattern = {on 3pt off 4pt},
]

\pgfdeclareplotmark{starmark}{
    \node[star,star point ratio=2.25,minimum size=6pt,
          inner sep=0pt,draw=black,solid,fill=red] {};
}

\def\algcommsymb{$\#$ }
\algrenewcommand\algorithmiccomment[1]{ \hfill{\color{Gray} \algcommsymb \textit{#1}}}
\NewDocumentCommand{\LineComment}{m}{\item[] {\color{Gray} \algcommsymb \textit{#1}}}

\let\gets\leftarrow
\algdef{SE}[DOWHILE]{Do}{doWhile}{\algorithmicdo}[1]
{\algorithmicwhile\ #1}%

\newcommand{\BreakIf}{\State \textbf{break if} }

\DeclareSIUnit{\sidb}{%
	\text{dB}
}

\def\m{\si{\meter}}
\def\dB{\si{\sidb}}
\let\db\dB
\def\cm{\si{\centi\meter}}
\def\ms{\si{\milli\second}}
\usetikzlibrary{positioning,calc}
\usetikzlibrary{decorations.pathreplacing}
\usetikzlibrary{shapes,shapes.geometric,shapes.arrows,decorations.pathmorphing}
\usetikzlibrary{matrix,chains,scopes,positioning,arrows,fit,backgrounds}
\usetikzlibrary{decorations.markings}
\usetikzlibrary{pgfplots.statistics, pgfplots.colorbrewer} 

\newlength{\minwidth}
\tikzstyle{basicblock} = [rectangle, minimum width=\minwidth, minimum height=1cm, text centered, draw=black]
\tikzstyle{input} = [basicblock, rounded corners, fill=blue!10]
\tikzstyle{output} = [basicblock, rounded corners, minimum width=2cm, fill=green!10]
\tikzstyle{minprocess} = [basicblock, fill=orange!10]
\tikzstyle{filtprocess} = [basicblock, fill=red!10]
\tikzstyle{decision} = [basicblock, chamfered rectangle, fill=yellow!80!red!10!white]

\tikzstyle{superinput} = [input, fill=blue!20]
\tikzstyle{superminprocess} = [minprocess, fill=orange!20]
\tikzstyle{superfiltprocess} = [filtprocess, fill=red!20]
\tikzstyle{superoutput} = [output, fill=green!20]

\tikzstyle{arrow} = [thick,->,>=stealth]

\tikzstyle{wavefront} = [%
    draw,thick,densely dotted,decorate,decoration={%
        markings, mark = between positions 0 and 1 step 2mm with {
        \draw[-,#1] (-0.03,0.06) -- (0,0) -- (-0.03,-0.06);
        }
    }
]
        
\tikzstyle{reflection} = [solid,thin,wavefront={#1}]
\input{setup/common/c0.my_accents.tex}

\newcommand{\pts}[1]{{\udel{#1}}}
\newcommand{\bts}[1]{\udel[{[}]{#1}[{]}]}
\newcommand{\cts}[1]{\udel[{\{}]{#1}[{\}}]}
\newcommand{\abs}[1]{\udel[|]{#1}[|]}
\newcommand{\norm}[1]{\udel[\lVert]{#1}[\rVert]}
\newsavebox{\jcsboxA}
\newsavebox{\jcsboxB}
\makeatletter
\newlength{\udel@lenA}
\newlength{\udel@lenB}
\NewDocumentCommand{\udel}{O{(} m O{)}}{%
	\left#1 #2 \right#3
}%
\makeatother

\input{setup/common/c2.my_matrix_operators.tex}
\NewDocumentCommand{\SubSize}{m m}{\IfValueT{#1}{_{\scalebox{0.7}{$ \sz{#1\!}{\!\IfValueTF{#2}{#2}{#1}} $}}}}

\newlength{\SMtblimit}
\NewDocumentEnvironment{smatrix}{O{\SMtblimit}}{
	\if@display
		\setlength{\arraycolsep}{3pt}%
		\setlength{\cellspacetoplimit}{#1}%
		\setlength{\cellspacebottomlimit}{#1}%
	\fi
	~\begin{matrix} }{
	\end{matrix}~
}
\NewDocumentEnvironment{sbmatrix}{O{\SMtblimit}}{
	\setlength{\arraycolsep}{3pt}%
	\setlength{\cellspacetoplimit}{#1}%
	\setlength{\cellspacebottomlimit}{#1}%
	\left[~\begin{matrix} }{
	\end{matrix}~\right]
}
\NewDocumentEnvironment{scmatrix}{O{\SMtblimit}}{
	\setlength{\arraycolsep}{3pt}%
	\setlength{\cellspacetoplimit}{#1}%
	\setlength{\cellspacebottomlimit}{#1}%
	\left\{~\begin{matrix} }{
	\end{matrix}~\right\}
}
\NewDocumentCommand{\otup}{O{}m o m o o}{%
	{%
		\begin{sbmatrix}%
			#2#1 & \cdots#1 & \IfValueT{#3}{#3#1 & \cdots#1 & } #4
		\end{sbmatrix}\,%
	}\SubSize{#5}{#6}%
}%
\NewDocumentCommand{\ovtup}{m o m o o}{
	{%
		\begin{sbmatrix}%
			#1 \\
			\vdots \\
			\IfValueT{#2}{%
				#2 \\
				\vdots \\
			}%
			#3
		\end{sbmatrix}\,%
	}\SubSize{#4}{#5}%
}%
\NewDocumentCommand{\pair}{m m o o}{%
	{%
		\begin{sbmatrix}%
			#1 & #2
		\end{sbmatrix}\,%
	}\SubSize{#3}{#4}%
}%
\NewDocumentCommand{\vpair}{m m o o}{%
	{%
		\begin{sbmatrix}%
			#1 \\[0.5em]
			#2
		\end{sbmatrix}\,%
	}\SubSize{#3}{#4}%
}%
\def\setsep{}
\newcommand{\setprocessor}[1]{
	\setsep\IfEmpty{#1}{\cdots}{#1}
	\global\def\setsep{\,,&}
}
\newcommand{\vsetprocessor}[1]{
	\setsep\IfEmpty{#1}{\vdots}{#1}
	\global\def\setsep{\\}
}

\NewDocumentCommand{\tup}{ >{\SplitList{,}}m o o }{
    \global\def\setsep{}
	\begin{sbmatrix}%
		\ProcessList{#1}{\setprocessor}
	\end{sbmatrix}\SubSize{#2}{#3}
	\global\def\setsep{}
}
\NewDocumentCommand{\vtup}{ >{\SplitList{,}}m o o }{
    \global\def\setsep{}
	\begin{sbmatrix}%
		\ProcessList{#1}{\vsetprocessor}
	\end{sbmatrix}\SubSize{#2}{#3}
	\global\def\setsep{}
}

\DeclareDocumentCommand{\added}{o m}{%
    {
    \color{LimeGreen} \IfValueT{#1}{\textit{[#1]} }%
    #2}%
}
\DeclareDocumentCommand{\removed}{o m}{%
    {
    \color{Red} \IfValueT{#1}{\textit{[#1]} }%
    \sout{#2}}%
}
\DeclareDocumentCommand{\changed}{o m m}{%
    {
    \color{Orange} \IfValueT{#1}{\textit{[#1]} }\textit{\sout{#2}}%
    #3}%
}

\def\todo{\color{Green}}

\makeatletter
\newcommand{\oset}[2]{{\mathpalette\o@set{{#1}{#2}}}}
\newcommand{\o@set}[2]{\o@@set{#1}#2}
\newcommand{\o@@set}[3]{%
  \vbox{\offinterlineskip
    \ialign{\hfil##\hfil\cr
      $\m@th\o@set@demote{#1}#2$\cr
      \noalign{\vskip0.2pt}
      $\m@th#1#3$\cr
    }%
  }%
}
\newcommand{\o@set@demote}[1]{%
  \ifx#1\displaystyle\scriptstyle\else
  \ifx#1\textstyle\scriptstyle\else
  \scriptscriptstyle\fi\fi
}
\makeatother

\MakeSymbol{\x}{\symsfup{x}}
\MakeSymbol{\y}{\symsfup{y}}
\MakeSymbol{\z}{\symsfup{z}}
\MakeSymbol{\w}{\omega}
\MakeSymbol{\an}{\symup{an}}
\MakeSymbol{\t}{\theta}
\MakeSymbol{\td}{\theta_{\mrm{d}}}
\MakeSymbol{\tB}{\theta_{\mrm{B}}}
\MakeSymbol{\Nu}{\mathcal{V}}
\MakeSymbol{\rn}{\symup{rn}}
\MakeSymbol{\+}{{}^{+}}
\MakeSymbol{\-}{{}^{-}}
\MakeSymbol{\Agemo}{\text{\rotatebox[origin=c]{180}{\ensuremath{\Omega}}}}

\MakeSymbol{\tD}{\tilde{D}}
\DeclareMathOperator{\dsrf}{DSRF}
\DeclareMathOperator{\isrf}{ISRF}
\DeclareMathOperator{\gsir}{gSIR}
\DeclareMathOperator{\gsnr}{gSNR}
\DeclareMathOperator{\df}{DF}
\DeclareMathOperator{\wng}{WNG}
\makeatletter
\NewDocumentCommand{\@OneIdx}{m m O{;} m}{
    #1_{\IfValueT{#2}{#2#3} {#4}}
}
\NewDocumentCommand{\@TwoIdx}{s m m m m}{
    #2_{\IfValueT{#3}{#3;}(#4\IfValueT{#3}{_{#3}}\IfBooleanT{#1}{'},#5\IfValueT{#3}{_{#3}}\IfBooleanT{#1}{'})}
}
\NewDocumentCommand{\@FourIdx}{m m m m m m}{
    #1_{\IfValueT{#2}{#2;}(#3\IfValueT{#2}{_{#2}},#4\IfValueT{#2}{_{#2}});(#5\IfValueT{#2}{_{#2}},#6\IfValueT{#2}{_{#2}})}
}

\NewDocumentCommand{\dx}{g}{
    \@OneIdx{\delta}{#1}{\x}
}
\NewDocumentCommand{\dy}{g}{
    \@OneIdx{\delta}{#1}{\y}
}
\NewDocumentCommand{\Mx}{g}{
    \@OneIdx{M}{#1}[,]{\x}
}
\NewDocumentCommand{\My}{g}{
    \@OneIdx{M}{#1}[,]{\y}
}

\DeclareDocumentCommand{\P}{g O{m}}{
    \@OneIdx{P}{#1}{#2}
}
\DeclareDocumentCommand{\S}{g O{m}}{
    \@OneIdx{S}{#1}{#2}
}
\DeclareDocumentCommand{\R}{g O{m}}{
    \@OneIdx{r}{#1}{#2}
} 
\DeclareDocumentCommand{\p}{g O{m}}{
    \@OneIdx{\psi}{#1}{#2}
}
\DeclareDocumentCommand{\D}{g O{m}}{
    \@OneIdx{D}{#1}{#2}
}
\DeclareDocumentCommand{\T}{g O{m}}{
    \@OneIdx{T}{#1}{#2}
}
\LetLtxMacro{\d}{\udot}
\DeclareDocumentCommand{\d}{g O{m}}{
    \@OneIdx{d}{#1}{#2}
}
\makeatother

\makeatletter
\DeclareDocumentCommand{\bv}{m O{n} g}{
	\@my{#2}{\symbfup{#1}}\IfValueT{#3}{_{#3}}
}
\DeclareDocumentCommand{\BV}{m g}{
    \symbfsfup{#1}\IfValueT{#2}{_{#2}}
}
\DeclareDocumentCommand{\newbv}{s o m}{
    \IfBooleanTF{#1}{
        \IfValueTF{#2}{
            \@namedef{bv#2}{\BV{#3}}
        }{
            \@namedef{bv#3}{\BV{#3}}
        }
    }{
        \IfValueTF{#2}{
            \@namedef{bv#2}{\bv{#3}}
        }{
            \@namedef{bv#3}{\bv{#3}}
        }
    }
}
\@namedef{:}{\scalebox{1.01}{\textbf{:}}}
\makeatother

\newbv{y}
\newbv{x}
\newbv{u}
\newbv{p}
\newbv{v}
\newbv[ga]{\upgamma}
\newbv[eta]{\upeta}
\newbv[ze]{\upzeta}
\newbv[mu]{\upmu}
\newbv[si]{\upsigma}
\newbv[vs]{\upvarsigma}
\newbv{a}
\newbv{b}
\newbv{t}
\newbv{z}
\newbv{j}
\newbv{w}
\newbv{q}
\newbv{p}
\newbv{A}

\newbv{I}
\let\Id\bvI
\newbv[Ga]{\Gamma}

\newbv[Xi]{\Xi}

\newbv{h}
\newbv{C}
\newbv{i}
\newbv{d}
\newbv{D}
\newbv{T}
\newbv[La]{\Lambda}
\newbv{g}
\newbv[up]{\upsilon}
\newbv{Q}
\newbv[Si]{\Sigma}
\newbv[ep]{\epsilon}
\newbv{f}
\newbv{P}
\newbv{n}

\def\lag{\mathcal{L}}
\def\sigs{\mathcal{S}}
\def\condition{\mathcal{C}}
\NewDocumentCommand{\std}{s m}{%
    \IfBooleanTF{#1}{\hat{\sigma}}{\sigma}_{#2}%
}
\NewDocumentCommand{\var}{s m}{%
	\IfBooleanTF{#1}{\std*{#2}}{\std{#2}}^2%
}
\NewDocumentCommand{\Corr}{s m}{%
    \IfBooleanTF{#1}{\bar{\bv{R}}}{\bv{R}}_{#2}%
}

\NewDocumentCommand{\corr}{s m}{%
	\IfBooleanTF{#1}{\hat{\bv{r}}}{\bv{r}}_{#2}%
}

\DeclareMathOperator{\mvdr}{MVDR}
\DeclareMathOperator{\lcmv}{LCMV}
\DeclareMathOperator{\lcmp}{LCMP}
\DeclareMathOperator{\music}{MUSIC}
\DeclareMathOperator{\wmusic}{wMUSIC}
\DeclareMathOperator{\filt}{f}
\MakeSymbol{\Tsixty}{%
    \mathrm{T}_{60}
}

\let\mrm\symup
\def\mtimes{\mathbin{\scalebox{0.7}{$\times$}}}
\NewDocumentCommand{\sz}{s m m}{%
	\IfBooleanTF{#1}{%
		\bts{{#2} \mtimes {#3}}%
	}{%
		{#2} \mtimes {#3}%
	}%
}

\NewDocumentCommand{\el}{m g O{m} o}{%
	\bts{#1}_{%
		\IfValueTF{#4}{%
			(#3\IfValueT{#2}{_{#2}},#4\IfValueT{#2}{_{#2}})%
		}{%
			#3\IfValueT{#2}{_{#2}}%
		}        
	}
}

\NewDocumentCommand{\FT}{m O{(\w)}}{%
	\symcal{F}\cts{#1}#2%
}
\NewDocumentCommand{\IFT}{m O{(t)}}{%
	\symcal{F}^{-1}\cts{#1}#2%
}

\NewDocumentCommand{\RFT}{m O{(\w)}}{%
	\symcal{S}\cts{#1}#2%
}
\NewDocumentCommand{\IRFT}{m O{(t)}}{%
	\symcal{S}^{-1}\cts{#1}#2%
}
\DeclareDocumentCommand{\S}{}{%
	\symcal{S}
}

\NewDocumentCommand{\STFT}{m O{[l,k]}}{%
	\symbb{F}\cts{#1}#2%
}
\NewDocumentCommand{\ISTFT}{m O{[n]}}{%
	\symbb{F}^{-1}\cts{#1}#2%
}

\NewDocumentCommand{\SSBT}{m O{[l,k]}}{%
	\symbb{S}\cts{#1}#2%
}
\NewDocumentCommand{\ISSBT}{m O{[n]}}{%
	\symbb{S}^{-1}\cts{#1}#2%
}

\NewDocumentCommand{\fnorm}{m}{%
	\norm{#1}_2%
}

\NewDocumentCommand{\opt}{s}{%
	\IfBooleanTF{#1}{^\dagger}{^\star}
}
\NewDocumentCommand{\cost}{s g}{%
    \IfBooleanTF{#1}{{F}}{\mathscr{F}}\IfValueT{#2}{\pts{#2}}%
}

\NewDocumentEnvironment{subalign}{g}{%
	\subequations%
		\IfValueT{#1}{\label{#1}}%
		\allowdisplaybreaks%
		\align%
	}{%
		\endalign%
	\endsubequations%
}%
\NewDocumentEnvironment{subgather}{g}{%
	\subequations%
		\IfValueT{#1}{\label{#1}}%
		\allowdisplaybreaks%
		\gather%
	}{%
		\endgather%
	\endsubequations%
}%
\NewDocumentEnvironment{equations}{g}{%
	\equation%
		\IfValueT{#1}{\label{#1}}%
		\allowdisplaybreaks%
		\aligned%
	}{%
		\endaligned%
	\endequation%
}

\NewDocumentEnvironment{proposition}{m}{%
	\noindent\textbf{Proposition:} {\itshape #1}
	
	\noindent\textbf{Proof:}\itshape
}{

	\noindent\bsquare
}

\NewDocumentEnvironment{example}{}{%
	\noindent\textbf{Example:}
}{

	\noindent\bsquare
}

\newcounter{smethod}
\newcounter{method}
\NewDocumentEnvironment{method}{s}{%
    \IfBooleanT{#1}{\stepcounter{method}}%
    \stepcounter{smethod}%
	{\noindent\textbf{Method \Alph{method}.\arabic{smethod}:}}
 
}{
    \vspace{-1em}\newline%
	\noindent\bsquare%
 
}

\newtheoremstyle{break}
{9pt}
{9pt}
{}
{}
{\bfseries}
{.}
{ }
{}
\theoremstyle{break}

\newtheorem{thm}{Theorem}
\crefname{thm}{Theorem}{Theorems}

\def\square{
	\tikz\fill[scale=0.25](0,0) -- (1,0) -- (1,1) -- (0,1) -- cycle;}
\newcommand{\tmark}{%
	{{\unskip\nobreak\vadjust{\nobreak}\hskip0pt\penalty50 \space
			\vadjust{}\nobreak\hfil\square%
			\clubpenalty=0 \widowpenalty=0 \brokenpenalty=0
			\parfillskip=0pt \finalhyphendemerits=0 \par
			\penalty 10000 \parskip=0pt\noindent}}\ignorespaces} 

\NewDocumentEnvironment{theorem}{m O{}}{
	\begin{thm}{\itshape #1}#2
		
		}{
			\newline\tmark
	\end{thm}
}

\begin{document}

\title{Robust DoA and Noise Covariance Matrix joint estimation for beamforming}

\NewDocumentCommand{\orcid}{m}{%
    $^{\orcidlink{#1}}$%
}

\author{%
    Vitor G. P. Curtarelli\orcid{0009-0009-3996-5452}, Stephan Paul\orcid{0000-0001-8181-1048}, and Anderson W. Spengler\orcid{0000-0001-5204-2040}
    \thanks{Manuscript Info: Corresponding author: Vitor G. P. Curtarelli.}%
    \thanks{V. G. P. Curtarelli and A. W. Spengler are with the Electrical and Engineering Department at Universidade Federal de Santa Catarina (UFSC), Florianópolis, SC, Brazil (emails: \url{vitor.curtarelli@posgrad.ufsc.br}, \url{anderson.spengler@ufsc.br}).}%
    \thanks{S. Paul is with the Mechanical Engineering Department at Universidade Federal de Santa Catarina (UFSC), Florianópolis, SC, Brazil (email: \url{stephan.paul@ufsc.br}).}%
} 

\markboth{IEEE TRANSACTIONS ON SIGNAL PROCESSING, Vol. X, 20YY}%
{Curtarelli et al.: DoA and NCM joint estimation}


\maketitle

\begin{abstract}	
	Robust beamforming in adverse acoustic environments has increasingly relied on accurate estimation of the Direction-of-Arrival (DoA) and the Noise Covariance Matrix (NCM). However, conventional approaches often face a trade-off between computationally prohibitive searches and sensitivity to reverberation. To address this, we propose a computationally efficient framework for the joint estimation of DoA and NCM. Unlike existing iterative methods that rely on exhaustive optimization, we derive a quasi-linear solution NCM estimator based on the Lagrange multiplier method, significantly reducing the algorithmic complexity. Furthermore, we introduce a DoA estimation technique designed to operate across the frequency spectrum, ensuring robustness in non-ideal, reverberant scenarios. Comparative simulations demonstrate that the proposed method outperforms classical subspace techniques, such as MUSIC, yielding lower angular errors in a wide range of real-life-like conditions. In signal enhancement tasks, our framework exhibits superior interference cancellation and preservation of the desired source, offering a practical and robust alternative for real-world beamforming applications.
\end{abstract}

\begin{IEEEkeywords}
	Noise covariance matrix, direction of arrival estimation, beamforming, microphone array processing.
\end{IEEEkeywords}
\section{Introduction}

Multi-sensor signal acquisition has become a fundamental pillar of modern signal processing, enabling precise analysis and enhancing output quality across diverse domains \cite{wang_overview_2019}. Sensor array-based processing techniques find application in scenarios ranging from assistive hearing devices \cite{jansen_benefit_2024} and smart home technologies \cite{haeb-umbach_speech_2019}, to radar systems \cite{aubry_geometric_2018,salari_joint_2019}, wireless communications \cite{diouf_usrpbased_2021,tuando-hong_signal_2004}, and biomedical instrumentation \cite{gaydecki_real_2000,qaisar_advances_2023}. In such settings, spatial filtering and signal enhancement are often essential, and hinge on accurate statistical models of the signal and its surrounding noise field. Among these, the estimation of signal characteristics --- such as covariance structures and source positioning --- plays a crucial role \cite{zoubir_robust_2012}.

A key tool in this context is the signal's covariance matrix (CM), characterizing the temporal-spatial relationships between received signals. While widely adopted, this approach can be severely affected by finite sample effects, noise, and rapidly changing environments \cite{dunik_noise_2017}. In these cases, relying directly on the observed signal's CM may result in suboptimal or biased signal enhancement performance \cite{muhammad_covariance_2022}.

An alternative lies in modeling the noise through a dedicated Noise Covariance Matrix (NCM) \cite{gres_subspacebased_2025,esfandiari_noise_2025}, isolating undesired components from the whole, and providing a more robust foundation for tasks such as beamforming, interference suppression, and source separation \cite{kjems_maximum_2012, dunik_autocovariance_2017}. Nonetheless, estimating the NCM presents challenges, such as the differentiation of signal and noisy sources, sensitivity to spatial correlations, and degradation under low-SNR conditions \cite{dunik_noise_2017}.

In many scenarios, NCM estimation is inherently linked to the estimation of the Direction-of-Arrival (DoA) of one or more interfering sources \cite{salari_joint_2019,barthelme_doa_2021}. However, traditional DoA estimation methods often rely on exhaustive search techniques, which are computationally expensive and sensitive to initial conditions \cite{stockle_doa_2015,yan_lowcomplexity_2013}.

In this work, we propose a novel method for jointly estimating the NCM of a contaminated sound field and the DoA of a directional interfering source. Building upon the compact NCM modeling approach in \cite{moore_compact_2022}, our method introduces a broadband cost function, extending the model to include the estimation of an interfering source's DoA, along with an improved mathematical framework that allows for quasi-linear estimation of the modeled signals' variances. This leads to a significant reduction in computational complexity while improving estimation accuracy and filtering performance. 

Importantly, our results show that the proposed DoA estimator outperforms traditional approaches such as MUSIC both in accuracy and precision, while enabling more effective beamforming when integrated into a full signal enhancement pipeline. These results are validated across almost 1500 combinations of environment parameters, demonstrating the proposed approach's robustness.


%
\section{Signal model}
\label{sec:signal_model}

We consider a sensor array (of any geometry) comprised of $M$ omnidirectional sensors within a reverberant environment populated by both desired and contaminating sources. We assume this environment to be stationary both spatially and for its statistical characteristics; these constraints are primarily for ease of notation and algebra.

The $m$-th observed signal in the time-domain is
\begin{equation}
	\label{eq:sec2:definition_observed-signal_time-domain}
	y_m(t) = s_m(t) + n_m(t) + \sum_{i=1}^{I} w_{m,i}(t) + v_m(t),
\end{equation}
where: $s_m(t)$ is the reverberant desired signal; $n_m(t)$ and $w_{m,i}(t)$ are reverberant undesired signals, with $n_m(t)$ being the dominant among these (called the interfering signal), and $w_{m,i}(t)$ originating from other (less impactful) undesired sources; and $v_m(t)$ is uncorrelated Gaussian noise, modeling thermal noise. These signals all refer to the $m$-th sensor within the sensor array.

Transforming all signals into the (discrete) time-frequency domain (with $l$ representing the time-frequency window index, and $k$ the frequency bin), and decomposing the $s_m(t)$ and $n_m(t)$ into a sum of planar waves \cite{moore_compact_2022}, we have the modeled observed signal $\tilde{y}_m[l,k]$
\begin{equations}{eq:sec2:model_hat-y_mlk}
	\tilde{y}_m[l,k] 
	& \approx y_m[l,k] \\
	& = \sum_{j_s=1}^{J_s} s_{m,j_s}[l,k] + \sum_{j_p=1}^{J_p} n_{m,j_p}[l,k] \\
	& + \sum_{i=1}^{I} w_{m,i}[l,k] + v_m[l,k],
\end{equations}
where $\tilde{y}_m[l,k]$ is the approximate model of $y_m[l,k]$, under the presented assumptions. Through the text, the tilde accent will denote a model or approximation for the accented variable.

Now, three modeling assumptions are taken:
\begin{enumerate}
    \item Each planar wave decomposition is dominated by a single plane-wave \cite{yilmaz_blind_2004} ($j_s = j_p = 1$), this being the direct-path wave between source and sensor;
    \item This direct-path component can be written in terms of the relative frequency response (RFR) between the reference ($m=1$) and the $m$-th sensors;
    \item All non-direct-path components of $s_m[l,k]$ and $n_m[l,k]$ ($j_s \geq 2$ and $j_p \geq 2$), and all non-dominant undesired sources $w_{m,i}[l,k]$, can be modeled by $\gamma_m[l,k]$.
\end{enumerate}

With these, the modeled signal from \cref{eq:sec2:model_hat-y_mlk} becomes
\begin{equations}
	\tilde{y}_m[l,k] = d_m[k] x_1[l,k] + b_m[k] p_1[l,k] + \gamma_m[l,k] + v_m[l,k],
\end{equations}
in which: $x_1[l,k]$ and $p_1[l,k]$ are (respectively) the desired and interfering signal's first planar-wave at the reference sensor; $\gamma_m[l,k]$ is the previously defined contaminating signal,
\begin{equation}
    \gamma_m[l,k] = \sum_{j_s=2}^{J_s} s_{m,j_s}[l,k] + \sum_{j_p=2}^{J_p} n_{m,j_p}[l,k] + \sum_{i=1}^{I} w_{m,i}[l,k];
\end{equation}
and $d_m[k]$ and $b_m[k]$ are (respectively) the desired and interfering signal's reference-to-$m$-th RFRs, given by
\begin{subgather}
	\label{eq:sec2:definition_steering-vector_dmk}
	d_m[k] = \e^{-\j 2\pi \frac{k f_0}{K\cdot c} r_m \cos(\theta_d - \psi_m) \cos(\phi_d - \lambda_m)}, \\
    b_m[k] = \e^{-\j 2\pi \frac{k f_0}{K\cdot c} r_m \cos(\theta_b - \psi_m) \cos(\phi_b - \lambda_m)},
\end{subgather}
with $\e$ denoting the Neperian constant, $f_0$ is the sampling frequency, $K$ is the total number of frequency bins, and $c = 343\si{\meter/\second}$ is the wave speed. The spherical coordinates $(r_m,~\psi_m,~\lambda_m)$ are the $m$-th sensor's position (distance, azimuth, elevation), with $(\theta_d,~\phi_d)$ and $(\theta_b,~\phi_b$) being the desired and interfering sources' angular directions (azimuth, elevation). All these spherical measurements assume the reference sensor as the origin.

We let $\bvy[t][l,k] = \tr{\big[\,\tilde{y}_1[l,k],~\cdots\!,~\tilde{y}_M[l,k]\,\big]}$ be the column vectorization of the modeled observed signals ($\bvd[k]$, $\bvb[k]$, $\bvga{}[l,k]$, and $\bvv[l,k]$ are defined similarly), with $\tr{(\cdot)}$ denoting the transpose operator. We also define $\bveta[t][l,k]$ as the modeled global noise vector, encompassing all non-desired signals, such that
\begin{equation}
	\bveta[t][l,k] = \bvb[k] p_1[l,k] + \bvga{}[l,k] + \bvv{},[l,k],\label{eqs:sec2:modeled_vector_bvyh_with_bveta:eq1}
\end{equation}
from which
\begin{subalign}
	\bvy[t][l,k]
    & = \bvd[k] x_1[l,k] + \bvb[k] p_1[l,k] + \bvga{}[l,k] + \bvv{}[l,k]
	\label{eq:sec2:modeled_vector_bvyh} \\
    & = \bvd[k] x_1[l,k] + \bveta[t][l,k].
\end{subalign}

An example of a considered environment layout is in \cref{fig:sec2:room_layout}, with desired and contaminating sources, considering their direct and reflecting paths.

\colorlet{reflectionColor}{gray!50!black}
\begin{figure}[t]
	\centering
    \begin{tikzpicture}[scale=1.2]
    	\draw[rounded corners,fill=white, ultra thick] (0,0) rectangle (5,5);
    	\coordinate (sensorArray) at (1.5,1.5);
    	\coordinate (desSource) at (3.5,1.5);
    	\coordinate (intSource) at (4,3.5);
    	\coordinate (corSource1) at ($(sensorArray)+({cos(22.5+45*3)},{sin(22.5+45*3)})$);
    	\coordinate (corSource2) at ($(sensorArray)+({cos(22.5+45*0)},{sin(22.5+45*0)})$);
    	\begin{scope}[opacity=0.12]
    	    \draw[reflection=reflectionColor] (desSource) -- (2.5,0) -- (sensorArray);
        	\draw[reflection=reflectionColor] (desSource) -- (1,0) -- (0,0.6) -- (sensorArray);
        	\draw[reflection=reflectionColor] (desSource) -- (5,0.93) -- (2.55,0) -- (0,0.95)-- (sensorArray);
        	\draw[reflection=reflectionColor] (desSource) -- (5,3.65) -- (4,5) -- (sensorArray);
        	\draw[reflection=reflectionColor] (intSource) -- (5,3.88) -- (1.6,5) -- (0,4.45) -- (5,2.7) -- (sensorArray);
        	\draw[reflection=reflectionColor] (intSource) -- (2.3,5) -- (0,2.8) -- (sensorArray);
        	\draw[reflection=reflectionColor] (intSource) -- (5,0.8) -- (4.7,0) -- (2.83,5) -- (sensorArray);
        	\draw[reflection=reflectionColor] (intSource) -- (0,2.9) -- (5,2.1) -- (sensorArray);
        	\draw[reflection=reflectionColor] (corSource1) -- (0,1) -- (0.6,0) -- (sensorArray);
        	\draw[reflection=reflectionColor] (corSource2) -- (2,5) -- (sensorArray);
            \draw[reflection=reflectionColor] (corSource2) -- (0.6,5) -- (0,3.95) -- (sensorArray);
    	\end{scope}
    	\foreach \x in {0,1,...,8}{
    		\draw[very thick,yellow!40!black,rotate=218.66] (intSource)+(-3:0.3+0.3*\x) arc [start angle=-3, end angle=3,radius = 0.3+0.3*\x];
    		\draw[very thick,green!40!black,rotate=180] (desSource)+(-5:0.2+0.2*\x) arc [start angle=-5, end angle=5,radius = 0.2+0.2*\x];
    	}
    	\foreach \x in {0,1,...,7}{
    		\coordinate (corSource) at ($(sensorArray)+({cos(22.5+45*\x)},{sin(22.5+45*\x)})$); 
    		\begin{scope}[shift=(corSource)]
    			\node[blue!40!black,scale=2.2] at (0,0) {\pgfuseplotmark{diamond*}};
    			\foreach \y in {0,1,2,3}{
    				\draw[thick,blue!40!black,rotate=180+22.5+45*\x] (0,0)+(-10:0.2+0.12*\y) arc [start angle=-10,end angle=10,radius=0.2+0.12*\y];				
    			}
    		\end{scope}
    	}
    	\node[green!40!black, scale=4.0] at (desSource) {\pgfuseplotmark{pentagon*}};
    	\node[green!80!black, scale=3.3] at (desSource) {\pgfuseplotmark{pentagon*}};
    	\node[yellow!40!black, scale=4.0] at (intSource) {\pgfuseplotmark{triangle*}};
    	\node[yellow!80!black, scale=3.0] at (intSource) {\pgfuseplotmark{triangle*}};
        \begin{scope}[scale=0.065,shift=(sensorArray)]
            \draw[rounded corners, fill=gray!25!white!70!red,thick] (-4.5,-4.5) rectangle (4.5,4.5);
            \foreach\x in {0,1,2,3}{
                \foreach\y in {0,1,2,3}{
                    \draw[fill=gray!10!white!90!red,thick] (-3+2*\x,-3+2*\y) circle(18pt);
                }
            }
    	\end{scope}
    \end{tikzpicture}
	\caption{Example of room layout, with: desired (green pentagon), interfering (yellow triangle), and correlated (blue diamond) sources; direct-path (opaque, with the color of the respective source) and reverberations (translucent gray arrows); and a sensor array (red rectangle), with sensors represented by small circles.}
	\label{fig:sec2:room_layout}
\end{figure}
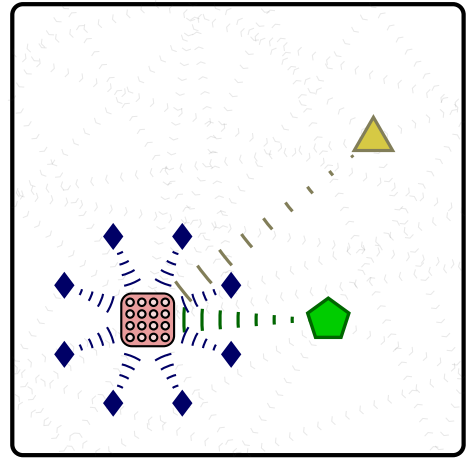 

\subsection{Observed signal covariance matrix modeling}
\label{subsec:sec2:covariance_matrix_modeling}
We write the observed signal's covariance matrix as $\Corr{\bvy}[k]$, and it is given by
\begin{equation}
	\Corr{\bvy}[k] = \expec[l]{\bvy[l,k] \he{\bvy}[l,k]},
\end{equation}
in which $\he{\{\cdot\}}$ represents the conjugate-transpose operation, and $\expec{\cdot}$ is the expectation operator (in this case, with respect to the window index $l$). We also define $\Corr{\bvy[t]}[k]$ as the modeled observed signal's CM, which --- given \cref{eq:sec2:modeled_vector_bvyh} --- can be expanded to
\begin{equation}
	\Corr{\bvy[t]} = \Corr*{\bvx} \var{x_1} + \Corr*{\bvp} \var{p_1} + \Corr*{\bvga} \var{\gamma_1} + \Corr*{\bvv} \var{v_1} + \epsilon\Id,
    \label{eq:sec2:modeled_CM_Rbvyh}
\end{equation}
where $\var{x_1}[k]$ is the variance of $x_1[l,k]$ and $\Corr*{\bvx}[k] = (\bvd \he{\bvd})[k]$ is its pseudo-normalized covariance matrix (these being denoted by the bar); and similarly for $\var{p_1}[k]$, $\var{\gamma_1}[k]$ and $\var{v_1}[k]$, as well as
$\Corr*{\bvp}[k] = (\bvb \he{\bvb})[k]$, $\Corr*{\bvga}[k]$ and $\Corr*{\bvv}[k]$. $\epsilon$ is a normalization factor, to ensure a minimal white noise is present on the modeled matrix. $\Corr*{\bvx}[k]$ is implicitly a function of $(\theta_d,\phi_d)$, and so is $\Corr*{\bvp}[k]$ (a function) of $\Theta_b=(\theta_b,\phi_b)$. While not explicit in \cref{eq:sec2:modeled_CM_Rbvyh}, all covariance matrices and variances (except $\epsilon$) are bin dependent, but not window dependent, due to the stationarity assumptions.

We now assume the following:
\begin{enumerate}
	\item The desired source's direction, as well as its steering vector $\bvd[k]$, are known;
	\item $\Corr*{\bvv}[l,k] = \Id$, since $\bvv[l,k]$ is uncorrelated white noise;
	\item $\bvga{}[l,k]$ is a spherically isotropic noise \cite{moore_compact_2022}, and thus $\Corr*{\bvga}[l,k]$ is known \cite{epain_spherical_2016};
	\item The DoA $\Theta_b$, which dictates $\bvb{}[l,k]$, is constant across the frequency spectrum.
\end{enumerate}

Although we take all (except $\Corr*{\bvp}[k]$) covariance matrices to be known, their respective variances are not. Therefore, we have 4K+2 unknowns in our problem: the variances for all signals at the reference sensor ($\var{x_1}$, $\var{u_1}$, $\var{\gamma_1}$ and $\var{v_1}$) for each frequency; and the DoA $\Theta_b$ which dictates the steering vector for the interfering signal.


Assuming these variances and $\Theta_b$ are known, the global noise covariance matrix $\Corr{\bveta}[k]$ can be estimated as $\Corr{\bveta[t]}[k]$, which (from \cref{eqs:sec2:modeled_vector_bvyh_with_bveta:eq1}) can be written as
\begin{equation}
	\label{eq:sec2:def_NCM_Corr-bveta}
	\Corr{\bveta[t]} = \Corr*{\bvp} \var{p_1} + \Corr*{\bvga} \var{\gamma_1} + \Corr*{\bvv} \var{v_1} + \epsilon \Id.
\end{equation}

We highlight that all matrices in \cref{eq:sec2:def_NCM_Corr-bveta} (including $\Corr{\bveta[t]} \equiv \Corr{\bveta[t]}[k]$) are frequency-dependent (similarly to all other covariance matrices), which was omitted for conciseness. 

\section{DoA and NCM estimation}
\label{sec:doa_ncm_estimation}

Given the desire to find the most appropriate values for the DoA $\Theta_b$ and for the variances, a minimization scheme must be carefully constructed. 

We define ${\bvsi{k} [k] = \bts{\var{x_1}[k],\var{p_1}[k],\var{\gamma_1}[k],\var{v_1}[k]}}$ as a vector of the unknown variances, and $\cost*{\Corr{\bvy[t]}(\bvsi{k},\Theta_b)[k]}$ as a frequency-dependent cost function, given by the Frobenius norm
\begin{equation}
	\label{eq:sec3:cost_function}
	\cost*{\Corr{\bvy[t]}(\bvsi{k},\Theta_b)}[k] = \fnorm{\Corr{\bvy[t]}[k] - \Corr{\bvy}[k]}^2,
\end{equation}
where this cost function is implicitly dependent on $\Theta_b$ and $\bvsi{k}$, as well as on the frequency bin $k$. By denoting ${R}_{\bvz;\bvj}$ as the $\bvj$-th element of $\Corr{\bvz}$ ($\bvj = [i,j]$), the cost function becomes
\begin{equations}{eq:sec3:cost_function_eachfreq}
	& \cost*\pts{\Corr{\bvy[t]}(\bvsi{k},\Theta_b)} [k] = \\
	& \sum_{\bvj} \abs{\bts{\sum_{\bvz\in\sigs} \bar{R}_{\bvz;\bvj}[k] \var{z_1}[k]} - \pts{{R}_{\bvy;\bvj}[k] - \epsilon I_{\bvj}}}^2,
\end{equations}
with $\sigs = \cts{\bvx,\bvp,\bvga,\bvv}$. For notation, $\Corr{\bvy,\epsilon} \equiv \Corr{\bvy} - \epsilon\Id$.

By splitting the absolute value's calculation into the squared sum of real and imaginary components, \cref{eq:sec3:cost_function_eachfreq} can be modified into \cref{eq:sec3:cost_function_longform}, where the superscripts $\re$ and $\im$ indicate the real and imaginary parts of the value, respectively.
\begin{equation}
	\begin{split}
		& \cost*{\Corr{\bvy[t]}(\bvsi{k},\Theta_b)} [k] \\
		& = \sum_{\bvj}
		\pts{\bts{\sum_{\bvz\in\sigs} \Re{\bar{R}_{\bvz;\bvj}}[k] \var{z_1}[k]} - \Re{R_{\bvy,\epsilon;\bvj}}[k]}^2 \\
		& + \sum_{\bvj}
		\pts{\bts{\sum_{\bvz\in\sigs} \Im{\bar{R}_{\bvz;\bvj}}[k] \var{z_1}[k]} - \Im{R_{\bvy,\epsilon;\bvj}}[k]}^2.
	\end{split}
	\label{eq:sec3:cost_function_longform}
\end{equation}

Given the frequency-dependent cost function $\cost*{\Corr{\bvy[t]}(\bvsi{k},\Theta_b)}[k]$, we now define our broadband cost function $\cost{\bvSi,\Theta_b}$ as
\begin{equation}
	\label{eq:sec3:cost_function_broadband}
	\cost{\bvSi,\Theta_b} = \sum_{k} \cost*{\Corr{\bvy[t]}(\bvsi{k}, \Theta_b)}[k],
\end{equation}
summing the frequency-dependent cost function over the spectrum; and $\bvSi$ containing the $4K$ variances for all frequency bins.

\subsection{Variances estimation}

Physically, the variances must be strictly non-negative, and therefore a minimization problem would need to satisfy this condition; that is, $\bvsi{k} \geq \bv{0}~\forall~k$. The subspace of $\re^4$ defined by this constraint will be called the \textit{feasible region}. An approach similar to the Lagrangian multiplier method \cite{rockafellar_lagrange_1993} will be employed to approach this problem, inserting the constraints into the minimization function. For such, we define the Lagrangian function $\lag(\bvSi,\Theta_b,\bvze,\bvmu)$ as
\begin{equation}
	\label{eq:sec3:lagrangian_function}
	\lag(\bvSi,\Theta_b,\bvze,\bvmu) = \cost{\bvSi,\Theta_b} - \tr{\bvze}\pts{\bvsi{k} - \bvmu},
\end{equation}
with $\bvmu = \bts{\mu_x^2,\mu_p^2,\mu_\gamma^2,\mu_v^2}$ being slack variables (given the inequality constraint), and $\bvze$ being the Lagrangian multiplier vector, both implicitly frequency-dependent. We denote $\pts{\bvSi\opt,\Theta_b\opt,\bvze\opt,\bvmu\opt}$ as the optimal solution that minimizes the Lagrangian, namely
\begin{equation}
	\label{eq:sec3:lagrangian_multiplier_eq}
	\bvSi\opt,\Theta_b\opt,\bvze\opt,\bvmu\opt = \argmin_{\bvSi,\Theta_b,\bvze,\bvmu} \lag(\bvSi,\Theta_b,\bvze,\bvmu).
\end{equation}

Since $\cost*{\Corr{\bvy[t]}(\bvsi{k},\Theta_b)}[k]$ for each frequency bin is independent from another bin in terms of their variances, taking the derivative of the Lagrangian w.r.t. (with respect to) any variance $\var{z_1}[k]$ yields
\begin{equation}
	\label{eq:sec3:partial_derivative_var-z}
	\pdv{\lag}{\var{z_1}[k]} = \sum_{\bvw\in\sigs} A_{\bvw;\bvz}[k] \var{w}[k] - A_{\bvy,\epsilon;\bvz}[k] - \zeta_{z}[k],
\end{equation}
where  $A_{\bvw;\bvz}$ is
\begin{equations}
	A_{\bvw;\bvz}[k]
	& = 2\sum_{\bvj} \Re{\bar{R}_{\bvw;\bvj}}[k] \Re{\bar{R}_{\bvz;\bvj}}[k] + \Im{\bar{R}_{\bvw;\bvj}}[k] \Im{\bar{R}_{\bvz;\bvj}}[k] ,
\end{equations}
$\bvw$ also acting as a placeholder variable ($\bvw \in \sigs \cup \{(\bvy,\epsilon)\}$). 
Note that the derivatives from \cref{eq:sec3:partial_derivative_var-z} are taken regarding $\var{z_1}[k]$ and not $\std{z_1}[k]$, as the latter would result in a cubic problem instead of a linear one, for which the direct solution would be ${\bvSi = \bv0}$ (corresponding to a local maximum).

Differentiating the Lagrangian w.r.t. $\zeta_{z}[k]$ and $\mu_{z}[k]$ yields
\begin{subgather}
	\label{eq:sec3:partial_derivative_xi-z}
	\pdv{\lag}{\zeta_{z}[k]} = \var{z_1}[k] - \mu_{z}[k], \\
	\label{eq:sec3:partial_derivative_mu-z}
	\pdv{\lag}{\mu_{z}[k]} = 2\zeta_{z}\mu_{z}.
\end{subgather}

The variances vector which minimizes the cost function from \cref{eq:sec3:cost_function_broadband} is one of the solutions where all derivatives in \cref{eq:sec3:partial_derivative_var-z,eq:sec3:partial_derivative_xi-z,eq:sec3:partial_derivative_mu-z} are zero. To find the physically meaningful solution (within the feasible region $\bvSi \geq \bv0$), the solution to the unconstrained variance minimization problem will be used to achieve the optimal variance vector for the constrained problem ($\bvSi \geq \bv0$).

\subsubsection{Unconstrained solution}
Since the Lagrangian w.r.t. any variance depends only on the variances in its own frequency bin (\cref{eq:sec3:partial_derivative_var-z}), our global minimization problem over all $4K$ variances becomes $K$ independent problems each with $4$ variables, and for each $k$ the (unconstrained) minimization can be written as
\begin{equation}
	\bvA[k] \bvsi{k} = \bvq[k],
\end{equation}
where
\begin{subgather}{subeqs:sec3:definition_bvA_bvq}
	\bvA[k] = \begin{bmatrix}
		A_{\bvx;\bvx}[k] & A_{\bvp;\bvx}[k] & A_{\bvga;\bvx}[k] & A_{\bvv;\bvx}[k] \\
		A_{\bvx;\bvp}[k] & A_{\bvp;\bvp}[k] & A_{\bvga;\bvp}[k] & A_{\bvv;\bvp}[k] \\
		A_{\bvx;\bvga}[k] & A_{\bvp;\bvga}[k] & A_{\bvga;\bvga}[k] & A_{\bvv;\bvga}[k] \\
		A_{\bvx;\bvv}[k] & A_{\bvp;\bvv}[k] & A_{\bvga;\bvv}[k] & A_{\bvv;\bvv}[k]
	\end{bmatrix}, \label{subeq:sec3:definition_bvA}\\
	\bvq[k] = \begin{bmatrix}
		A_{\bvy,\epsilon;\bvx}[k] + \zeta_{x}[k]\\
		A_{\bvy,\epsilon;\bvp}[k] + \zeta_{u}[k]\\
		A_{\bvy,\epsilon;\bvga}[k] + \zeta_{\gamma}[k]\\
		A_{\bvy,\epsilon;\bvv}[k] + \zeta_{v}[k]
	\end{bmatrix}, \label{subeq:sec3:definition_bvq}
\end{subgather}
for which the solution is
\begin{equation}
	\label{eq:sec3:solution_unconstrained-minim_bvsi}
	\bvsi{k}^{\star}[k] = \inv{\bvA}[k] \bvq[k].
\end{equation}

This result doesn't necessarily respect ${\bvsi{k} \geq \bv{0}}$.

\subsubsection{Constrained solution}
For a generic $z$, from \cref{eq:sec3:partial_derivative_mu-z} we have two options: either $\zeta_z[k] = 0$, meaning the constraint is inactive and the minimization is unrestricted in $z$; or $\mu_z[k] = 0$, implying ${\var{z_1}[k] = 0}$ (from \cref{eq:sec3:partial_derivative_xi-z}), meaning the constraint is active. The unconstrained problem's process can be adapted to find the restricted problem's solution through \cref{alg:active_inactive_analysis}; with its related proofs in \cref{thm:no-constraints_global_min_positive_values,thm:constrained_min_boundary}, from \cref{app:appB:proofs}.

\begin{algorithm}[!t]
	\caption{Active-inactive constraints analysis}
    \label{alg:active_inactive_analysis}
	\begin{algorithmic}[1]
        \State $\bvsi\opt \gets \argmin\limits_{\bvsi} \cost*{\bvsi}$ \Comment{Global minimum}
        \State $S \gets \{\bvsi\opt<0\}$ \Comment{Set of negative entry indices in $\bvsi\opt$}
        \For {$n = 1,\ldots,\abs{\{\bvsi\opt<0\}}$}
            \For {$c \in \mathbf{P}_{n}(S)$}
                \LineComment{$\mathbf{P}_n(S)$ is all $n$-element sets in the power-set of $S$}
                \LineComment{$c$ is a combination with $n$ active constraints}
                \State $\bvsi\opt \gets \argmin\limits_{\bvsi} \cost*{\bvsi}$ with active $c$-th constraints
                \State $\bvsi\opt* \gets \bvsi$ if $\cost*{\bvsi} < \cost*{\bvsi\opt*}$
            \EndFor
            \BreakIf $\bvsi\opt* \geq \bv0$
        \EndFor
	\end{algorithmic} 
\end{algorithm}

If the $z$-th constraint is active ($\var{z_1} = 0$), its respective entries from $\bvsi$ $\bvq[k]$, and rows and column from $\bvA[k]$, are unnecessary. Given there are $2^4=16$ combinations of active-inactive constraints, at most $16$ matrix inversions are sufficient to achieve the optimal constraint solution, with the worst case being with $\bvsi\opt* = \bv0$. Therefore, we say that a quasi-linear solution to the variances minimization can be achieved, not relying on iterative processes. This analysis is repeated for all $k$ frequency bins, achieving the $4K$ solutions for the variances.

The active-inactive analysis on $\bvsi{k}$ has to be repeated for each iteration of the minimization on $\Theta_b$, since for one direction the constraints may have to be active, but not for a different neighboring direction.

\subsection{DoA estimation}

The previous minimization scheme solved only for $\bvSi$, treating $\Theta_b$ as a constant. Since any element of $\bvA[k]$ and $\bvq[k]$ that corresponds to $\bvp[k]$ depends on $\Theta_b$, so do the achieved solutions. That is, for each direction $\Theta_b$, we can find a solution on $\bvSi$ which minimizes the error between $\Corr{\bvy[t]}[k]$ and $\Corr{\bvy}[k]$, within the feasible region defined by the positive-variance constraints; but this achieved solution isn't necessarily a global minimum, being necessary a minimization on $\Theta_b$.

By definition, $\bar{R}_{\bvp;\bvj}(\theta_b,\phi_b)$ depends on the relative position between sensors $i$ and $j$ ($\bvj = [i,j]$), and can be written as
\begin{equation}
	\bar{R}_{\bvp;\bvj}(\Theta_b,\phi_b) = \e^{-\j \frac{2\pi k f_0}{K c} r_{\bvj} \cos\pts{\Theta_b - \psi_{\bvj}} \cos\pts{\phi_b - \lambda_{\bvj}}},
\end{equation}
where $r_{\bvj}$, $\psi_{\bvj}$ and $\lambda_{\bvj}$ are the relative spherical coordinates (distance, azimuth, and elevation) between the $i$-th and $j$-th sensors. Since the term $\tr{\bvze}\pts{\bvsi{k} - \bvmu}$ doesn't depend on any of $\theta_b$ or $\phi_b$, the Lagrangian's (\cref{eq:sec3:lagrangian_function}) derivatives w.r.t. $\theta_b$ and $\phi_b$ are, respectively,
\begin{equation}
	\label{eq:sec3:pdv_lagrangian_theta}
	\pdv{\lag}{\theta_b} = -\frac{4\pi f_0}{K c}\sum_{\bvj} f_{1;\bvj}(\theta_b,\phi_b) G_{\bvj}(\theta_b,\phi_b),
\end{equation}
\begin{equation}
	\label{eq:sec3:pdv_lagrangian_phi}
	\pdv{\lag}{\phi_b}   = -\frac{4\pi f_0}{K c}\sum_{\bvj} f_{2;\bvj}(\theta_b,\phi_b) G_{\bvj}(\theta_b,\phi_b),
\end{equation}
with
\begin{subgather}{eqs:sec3:parameters_pdv_lagrangian_dirs}
	f_{1;\bvj}(\theta_b,\phi_b) = \sin\pts{\theta_b - \psi_{\bvj}} \cos\pts{\phi_b - \lambda_{\bvj}}, \\
	f_{2;\bvj}(\theta_b,\phi_b) = \cos\pts{\theta_b - \psi_{\bvj}} \sin\pts{\phi_b - \lambda_{\bvj}}, \\
	G_{\bvj}(\theta_b,\phi_b) = r_{\bvj} \sum_{k} k\var{p_1}[k] \Im{ \cts{\hat{R}^*_{\bvj}[k] \bar{R}_{\bvp;\bvj}(\theta_b,\phi_b)[k]} } ,
\end{subgather}
where $\hat{R}_{\bvj}[k]$ is the $\bvj$-th element of $(\Corr{\bvy[t]}[k] - \Corr{\bvy,\epsilon}[k])$, and $[\cdot]^*$ denotes the complex conjugate operation. Moving forward the $\nfrac{4\pi f_0}{c}$ term will be ignored, given it is a positive constant. We also consider the following:
\vspace*{-1em}
\begin{itemize}
	\item For any diagonal element, $r_{[i,i]} = 0$;
	\item For any element $-\bvj \equiv [j,i]$:
	\begin{itemize}
		\item[$\circ$] $r_{-\bvj} = r_{\bvj}$
		\item[$\circ$] $\psi_{-\bvj} = \psi_{\bvj} + \pi$
		\item[$\circ$] $\lambda_{-\bvj} = -\lambda_{\bvj}$
		\item[$\circ$] $G_{-\bvj}(\theta_b,\phi_b) = -G_{\bvj}(\theta_b,\phi_b)$
	\end{itemize}
\end{itemize}

With these properties, we have that
\begin{subgather}
	f_1(\theta_b,\phi_b,-\bvj) = -\sin(\theta_b - \psi_{\bvj}) \cos(\phi_b + \lambda_{\bvj}), \\
	f_2(\theta_b,\phi_b,-\bvj) = -\cos(\theta_b - \psi_{\bvj}) \sin(\phi_b + \lambda_{\bvj}).
\end{subgather}

By gathering the $\bvj$ and $-\bvj$ terms in the summations of \cref{eq:sec3:pdv_lagrangian_theta,eq:sec3:pdv_lagrangian_phi}, they simplify to
\begin{subgather}{eqs:sec3:pdv_cost_angles_simplified}
	\pdv{\lag}{\theta_b} = 2\cos\pts{\phi_b}\sum_{\substack{\bvj=[i,j]\\i<j}} \sin\pts{\theta_b - \psi_{\bvj}} \cos\pts{\lambda_{\bvj}} G_{\bvj}(\theta_b,\phi_b), \label{eqs:sec3:pdv_cost_angles_simplified:subeq1}\\
	\pdv{\lag}{\phi_b} = 2\sin\pts{\phi_b}\sum_{\substack{\bvj=[i,j]\\i<j}} \cos\pts{\theta_b - \psi_{\bvj}} \cos\pts{\lambda_{\bvj}} G_{\bvj}(\theta_b,\phi_b). \label{eqs:sec3:pdv_cost_angles_simplified:subeq2}
\end{subgather}

Both $\pdv{\lag}{\theta_b}$ and $\pdv{\lag}{\phi_b}$ are heavily non-linear, and a direct solution isn't available. However, as this is an unconstrained minimization problem, an arbitrarily accurate solution can be achieved through iterative or exhaustive approaches. We propose the use of the gradient descent technique, as both the cost function and its derivatives are available. Since it requires an initial guess, our minimization process can either use a multi-initialization step, or a previously estimated DoA (assuming spatially and/or statistically non-stationary systems).

\label{subsec:sec3:approximations_specific-array_considerations}

Although the established process is generic and works with any sensor configuration, simplifications and approximations can be explored, for the general case and for specific scenarios.

\subsubsection{General approximation}

From \cref{eqs:sec3:pdv_cost_angles_simplified}, we have that $\phi_b = \pm90\dg$ leads to a null in $\npdv{\lag}{\theta_b}$, and $\phi_b = 0\dg$ a null in $\npdv{\lag}{\phi_b}$. However, these nulls are due to the geometric construction and choice of variables, and unrelated to the DoA estimation procedure. We postulate it to be safe to ignore the leading $\phi_b$'s sine/cosine terms when calculating the derivative, as they can interfere and cause false results. Formally, we say that
\begin{subgather}
	\pdv{\lag}{\theta_b} \approx \sum_{\substack{\bvj=[i,j]\\i<j}} \sin\pts{\theta_b - \psi_{\bvj}} \cos\pts{\lambda_{\bvj}} G_{\bvj}(\theta_b,\phi_b), \\
	\pdv{\lag}{\phi_b} \approx \sum_{\substack{\bvj=[i,j]\\i<j}} \cos\pts{\theta_b - \psi_{\bvj}} \cos\pts{\lambda_{\bvj}} G_{\bvj}(\theta_b,\phi_b),
\end{subgather}
are sufficiently good approximations for the derivatives. In a case where $\phi_b = 0\dg$ or $\phi_b = 90\dg$ would minimize the cost function, they would also be roots in this new approximation.

\subsubsection{Planar and linear arrays}
\label{subsubsec:planar_arrays}
If one is working with a planar sensor array, it is easy to see that $\lambda_{\bvj} = \lambda_{\bvj'}~\forall~\bvj,\bvj'$, and therefore the $\cos(\lambda_{\bvj})$ term in both \cref{eqs:sec3:pdv_cost_angles_simplified:subeq1,eqs:sec3:pdv_cost_angles_simplified:subeq2} can be treated as a constant and ignored.

For the linear array case, one useful consideration is its steering vector's cylindrical symmetry. Given a direction $(\theta,\phi)$, there exists another direction $(\theta',0)$ such that their perceived response are identical. It is easy to see that
\begin{equation}
	\theta' = \arccos\pts{\cos(\theta) \cos(\phi)},
\end{equation}
achieves this result. In this scenario, we assume that $\phi_b = 0$, and the DoA estimation is reduced to a single variable. Furthermore, the azimuth for all sensors is constant as well, and we can safely assume $\psi_{\bvj} = 0 ~\forall~ \bvj$ (through a coordinate system rotation). Thus $\sin(\theta_b)$ is independent of $\bvj$ and can be factored and ignored. From these considerations, we get
\begin{equation}
	\pdv{\lag}{\theta_b} \approx \sum_{\substack{j=[i,j]\\i<j}} G(\theta_b,0,\bvj).
\end{equation}

\subsection{Minimization scheme}

Having an optimization pipeline on $\bvsi{k}$ and $\Theta_b$, the global minimization scheme is achieved through the steps in \cref{alg:global_minim_scheme}.

\begin{algorithm}[!htbp]
	\caption{Global minimization scheme}
    \label{alg:global_minim_scheme}
	\begin{algorithmic}[1]
        \State $\Theta_b \gets \Theta_{b;0}$ \Comment{Initial estimate}
        \Do
            \State $\bvSi\opt \gets \argmin\limits_{\bvSi} \lag({\bvSi,\Theta_b,\bvze,\bvmu})$ \Comment{ \cref{alg:active_inactive_analysis}} \vspace*{0.5em}
            \State $\Theta_b\opt \gets \argmin\limits_{\Theta_b}\lag({\bvSi,\Theta_b,\bvze,\bvmu})$ \Comment{Gradient descent}\vspace*{0.5em}
        \doWhile{$\pdv{\lag}{\theta_b},\pdv{\lag}{\phi_b} > \epsilon$} \Comment{Convergence condition} \vspace*{0.5em}
        \State $\Corr*{\bvp} \gets \bvb \he{\bvb}$
        \State $\Corr{\bveta[t]} \gets \Corr*{\bvp} \var{p_1} + \Corr*{\bvga} \var{\gamma_1} + \Corr*{\bvv} \var{v_1} + \epsilon \Id$ \Comment{\cref{eq:sec2:def_NCM_Corr-bveta}}
        \LineComment{$\Corr*{\bvp}$ and $\Corr{\bveta[t]}$ are constructed for every bin $k$}
	\end{algorithmic} 
\end{algorithm}

\subsection{Proposed model's features}
The proposed estimation method exhibits several noteworthy properties. Its main appeal is its ability to jointly estimate the NCM and an interfering source's DoA, improving signal enhancement performance. Its formulation is also inherently broadband (although applicable to narrowband signals), leveraging spectral information and improving robustness in reverberant or noisy conditions. Other useful features are its ability to estimate both azimuth and elevation angles, and its independence of a voice activation detector, common in NCM estimators.

The method's iterative nature enables source tracking by periodically updating the DoA, assuming the source position does not vary too rapidly. However, this iterative approach introduces some limitations, such as possible computational demand, depending on number of required steps. Its current form also isn't suitable for multi-source tracking.
\section{Filtering and Beamforming}
\label{sec:filtering_beamforming}

Given the observed signal $\bvy[l,k]$, a filter can be applied on it, leveraging the environmental information available (through $\bvd[k]$, $\bvb[h][k]$ and $\Corr*{\bveta}[k]$). This is done through a linear filter $\bvh[l,k]$, producing an estimate $\tilde{x}[l,k]$ that best approximates the desired signal at reference. This filtering process is defined by
\begin{equations}
	\tilde{x}[l,k] 
	& \approx x_1[l,k] \\
	& = \he{\bvh}[l,k] \bvy[l,k] \\
	& = \he{\bvh}[l,k] \bvd[k] x_1[l,k] + \he{\bvh}[l,k] \bveta[l,k].
\end{equations}

The desired signal estimation can be achieved through the constraint $\he{\bvh}[l,k] \bvd[k] = 1$ (this being called the distortionless constraint); and the reduction of the residual noise $\he{\bvh}\bveta$ is done through its direct minimization, possibly with additional constraints. Based on this framework, several beamformers can be designed, each utilizing different information about the signals and environment, and aiming for distinct optimization goals.

\subsection{NCM+DoA exploiting filters}

To exploit both (,) estimations obtained via \cref{sec:doa_ncm_estimation}, the linearly-constrained minimum variance (LCMV) beamformer can be used \cite{souden_study_2010}. In addition to the distortionless constraint, it allows the placement of beampattern nulls in directions of known interference. In our case, we aim to use the interfering source's estimated DoA ${\Theta}_b\opt$ and its steering vector $\bvb[k]$ to cancel $\bvp[l,k]$, the directional portion of the interfering signal $\bvn[l,k]$. The LCMV beamformer $\bvh{\lcmv}[k]$ is given by
\begin{equation}
	\bvh{\lcmv} = \inv{\Corr{\bveta[t]}} \bvC \inv*{\he{\bvC} \inv{\Corr{\bveta[t]}} \bvC} \bvi,
	\label{eq:sec4:beamformer_lcmv}
\end{equation}
with $\bvC[k]$ being a $\sz{M}{2}$ concatenation matrix $\bvC = \bts{ \bvd , \bvb }$, and $\bvi = \tr{[1, 0]}$ a $\sz{2}{1}$ vector.

An alternative approach that only partially utilizes the estimated information is the minimum-variance distortionless response (MVDR) beamformer \cite{souden_study_2010}, which preserves the desired signal while minimizing the global noise, without the null on $\Theta_b\opt$. The standard MVDR formulation for the beamformer, denoted $\bvh{\mvdr}[k]$, is
\begin{equation}
	\bvh{\mvdr} = \frac{ \Corr{\bveta[t]} \bvd{} }{ \he{\bvd} \Corr{\bveta[t]} \bvd{} }.
	\label{eq:sec4:beamformer_mvdr}
\end{equation}

\subsubsection{Full-framework outline}

\begin{figure}[t]
	\centering
	\begin{tikzpicture}[node distance=0.4cm]
    	\small\centering
    	\node (inDoa) [input,align=center] {Interfering\\DoA guess};
    	\node (inObs) [input,right=of inDoa,xshift=-0.3cm] {Observed signal};
    	\node (inDesDoa) [input,right=of inObs,xshift=-0.3cm,align=center]  {Desired\\signal's DoA};
    	\node (inLbl)[above right] at (inDoa.north west) {Inputs};
    	\node (minVar) [minprocess, below=of inDoa,yshift=-1cm,xshift=0.6cm] {Estimate variances};
    	\node (minLbl)[above right] at ([xshift=-1.2em]minVar.north west){Minimization};
    	\node (minDoa) [minprocess, below=of minVar] {Estimate DoA};
    	\node (minTest) [decision, below=of minDoa] {Converged?};
    	\node (minRP) [left= of minTest,xshift=0.2cm,label={[anchor=north,yshift=-0.5em,xshift=0.2em]No}] {};
    	\node (estNcm) [minprocess, below=of minTest] {Estimate NCM};
    	\node (calcFilt) [filtprocess] at ([xshift=-0.3cm]inDesDoa |- minDoa) {Calculate filter};
    	\node (filtLbl)[above right] at ([xshift=0.2ex]calcFilt.north west){Filtering};
    	\node (filtSig) [filtprocess, below=of calcFilt] {Filter signal};
    	\node (outSig) [output] at ([yshift=-2.4cm] filtSig|-estNcm) {Filtered signal};
    	\node (outDoa) [output,left=of outSig,xshift=0.3cm,align=center] {Interfering\\signal's DoA};
    	\node (outNcm) [output,left=of outDoa,xshift=0.3cm] {NCM};
    	\node (outLbl)[above right] at ([xshift=0.2ex]outNcm.north west){Outputs};
    	\begin{scope}[on background layer]
    		\node[superinput, fit={(inLbl) (inObs) (inDesDoa) (inDoa)},draw] (inSuper) {};
    		\node[superminprocess, fit={(minLbl) (minVar) (minDoa) (minTest) (minRP) (estNcm)},draw] (minSuper) {};
    		\node[superfiltprocess, fit={(filtLbl) (filtSig) (calcFilt)},draw] (filtSuper) {};
    		\node[superoutput,fit={(outLbl) (outSig) (outDoa) (outNcm)}, draw] (outSuper) {};
    	\end{scope}
    	\node (midTop) at ($(inDoa)!0.4!(minVar)$) {};
    	\node (midBot1) at ($(minSuper.south)!0.5!(outSuper.north)$) {};
    	\node (midBot2) at ($(filtSuper.south)!0.5!(outSuper.north)$) {};
    	\draw [arrow,densely dashed] (inDoa) |- ([xshift=-0.35cm]inObs |- midTop) |- ([yshift=0.2cm]minVar.east);
    	\draw [arrow,densely dashed] (inObs.south) |- (minVar.east);
    	\draw [arrow,densely dashed] (inDesDoa) |- ([xshift=0.35cm]inObs |- midTop) |- ([yshift=-0.2cm]minVar.east);
    	\draw [arrow,densely dotted] (minTest.south) -- (estNcm) node[pos=0.5,right] {Yes};
    	\draw [arrow,densely dotted] (minVar) -- (minDoa);
    	\draw [arrow,densely dotted] (minDoa) -- (minTest);
    	\draw [arrow,densely dotted] (minTest.west) -- (minRP.center) |- (minVar.west);
    	\draw [arrow,densely dashed] ([yshift=-0.2cm,xshift=0.35cm] inObs |- minVar) |- ([yshift=0.2cm]calcFilt.west);
    	\draw [arrow,densely dotted] (minDoa) -- (calcFilt.west);
    	\draw [arrow,densely dotted] (estNcm) -| ([yshift=-0.2cm,xshift=0.35cm] inObs |- calcFilt) -- ([yshift=-0.2cm] calcFilt.west);
    	\draw [arrow,densely dashed,preaction={draw=white, -, line width=3pt,shorten <=0.2em, shorten >=1em}] (inObs.south |- minVar.west) |- (filtSig.west);
    	\draw [arrow,densely dotted] (calcFilt) -- (filtSig);
    	\draw [arrow] (estNcm) |- (outNcm |- midBot1) -| ([xshift=0.3cm]outNcm.north);
    	\draw [arrow,preaction={draw=white, - , line width=3pt, shorten <=0.2em, shorten >=3em}] ([xshift=-0.5cm]outDoa |- minDoa) |- (outDoa |- midBot1) -- (outDoa);
    	\draw [arrow] (filtSig) |- (outSig.north |- midBot1) -| (outSig.north);
    \end{tikzpicture}
	\caption{Proposed NCM-DoA estimation technique + filtering scheme flowchart.}
	\label{fig:sec4:proposed_flowchart}
\end{figure}
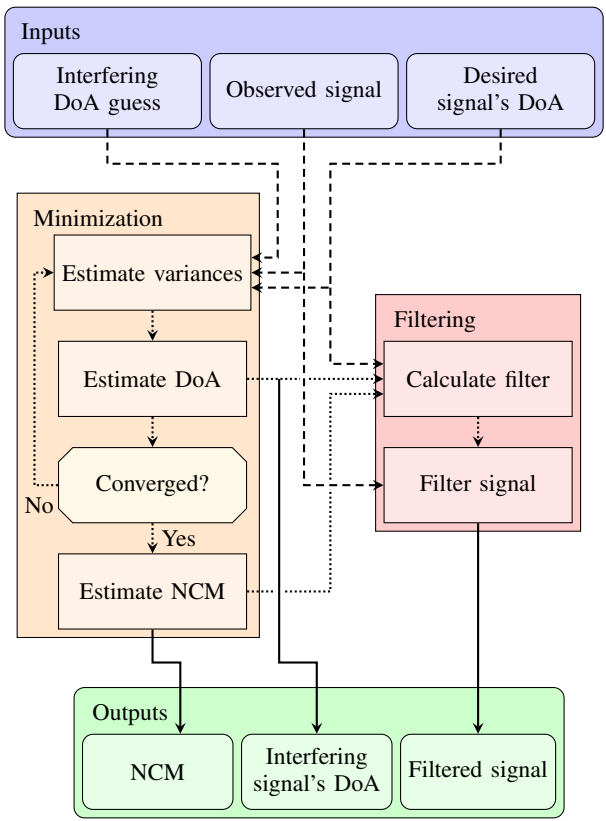

A diagram that presents the proposed technique's flowchart is in \cref{fig:sec4:proposed_flowchart}. Necessary input information is in blue, the minimization scheme --- a simplification from \cref{sec:doa_ncm_estimation} --- in yellow, the filtering process (with any beamformer that exploits the obtained NCM and/or DoA estimations) in red, and the output information in green. Input information flow is shown in dashed lines, internal logic flow in dotted lines, and output flow in solid lines.

\subsection{Straightforward filter}

A final option is adapting the LCMV beamformer to utilize the observed signal's CM $\Corr{\bvy}$ instead of the estimated NCM $\Corr{\bveta[t]}$; this being the Linearly-Constrained Minimum Power (LCMP) beamformer \cite{vergallo_processing_2012}. It is defined as
\begin{equation}
	\bvh{\lcmp} = \inv{\Corr{\bvy}} \bvC \inv*{\he{\bvC} \inv{\Corr{\bvy}} \bvC} \bvi.
	\label{eq:sec4:beamformer_lcmp}
\end{equation}

Since the LCMP beamformer uses the observed signal’s CM --- which can be estimated directly from the data --- it is compatible with external DoA estimation methods, such as MUSIC \cite{gupta_music_2015}. While the LCMV and LCMP beamformers are strictly equivalent in ideal conditions, it is known that LCMP designs are more sensitive to steering errors \cite{vergallo_processing_2012} and can lead to distortion on the desired signal.
\section{Simulations}

In this section, we present the different simulated acoustic scenarios, and compare the achieved models to the literature. We will first analyze the DoA estimation, and later the filtering process.

\subsection{Scenarios: sources, receivers and environment}

Simulated scenarios were used, considering an anechoic or slightly reverberant $5\m\times6\m\times3\m$ room. The sensor array's center is an uniform rectangular array with $16$ sensors in a $\sz{4}{4}$ grid, with intersensor distance of $\delta=2\si{\centi\meter}$ in both directions, centered at $(1.5\m,1.5\m,1\m)$. The array's center is the origin for the environment's coordinate system. Nonetheless, for simplicity the estimation process' measurements (sensor-to-sensor angles and distances, and estimated DoA) are relative to the reference sensor.


All sources are positioned at $0\dg$ elevation --- lying in the array's horizontal plane --- and therefore only $\theta_b$ needs to be estimated. The desired source's azimuth is fixed at $\theta_d = 0\dg$, and the correlated sources are positioned $1\m$ away from the origin, being 8 sources uniformly distributed in a circle. The approximations from \cref{subsec:sec3:approximations_specific-array_considerations} were also taken.

\begin{table}[!t]
	\centering
	\renewcommand{\arraystretch}{1.4}
	\caption{Simulation parameters, totaling $1458$ configurations in total.}
	\label{tab:sec5:simulation_parameters}
	\begin{tabular}{r l l l}
		Parameter & \multicolumn{3}{c}{Possible values} \\
		\hline \hline
		$\Tsixty$ & $0\ms$ & $500\ms$ & $800\ms$ \\
		$d_{x}$ & $50\cm$ & $150\cm$ & $300\cm$ \\
		$d_{p}$ & $50\cm$ & $150\cm$ & $300\cm$ \\
		SIR & $-10\db$ & $0\db $& $5\db$ \\
		SCR & $0\db$ & $5\db$ & $10\db$ \\
		$\theta_b$ & \multicolumn{3}{c}{$10\dg$~ $30\dg$~ $50\dg$~ $70\dg$~ $90\dg$~ $110\dg$} 
	\end{tabular}
\end{table}

The reverberation time $\Tsixty$, desired and interfering sources' distances $d_x$ and $d_p$ (relative to the sensor array's center), the interfering source's azimuth $\theta_b$, as well as Signal-to-Interference Ratio (SIR) and Signal-to-Correlated Ratio (SCR) are given in \cref{tab:sec5:simulation_parameters}, totaling $6\cdot3^5=1458$ parameter combinations. Signal power ratios are calculated between the desired signal's direct path component, and the respective reverberant contaminating signal. These possible parameter values were chosen to sweep a wide range of conditions, covering optimal and sub-optimal scenarios for both DoA estimation and filtering outcome.

The desired, interfering, and correlated sources' signals are respectively a male voice, a female voice, and music excerpts; these being obtained from the SMARD database \cite{nielsen_single_2014}. The room impulse responses (RIRs) between sources and sensors were generated using Habets' RIR generator \cite{habets_generating_2007,habets_room_2020}, where the sources and sensors are assumed to be omnidirectional. All signals and RIRs used were resampled to $16\si{\kilo\hertz}$, and the time-frequency signals are obtained through the STFT, with $K = 64$ bins, and Hamming windows with $50\%$ overlap. The regularization parameter $\epsilon$ (\cref{eq:sec2:modeled_CM_Rbvyh}) is set to $0.0001$, to ensure that a minimal white noise is considered.

\subsubsection{Benchmark algorithm}

The proposed DoA estimation method will be compared to 
the traditional MUSIC algorithm \cite{gupta_music_2015}. For the MUSIC implementation, the angular search space was discretized using an exhaustive search with a fineness of $1\dg$. To define a challenging test case, a scenario with a single dominant interfering source was simulated. The interferer's estimated DoA was determined by selecting the largest spectral peak from the MUSIC pseudospectrum, provided its angular separation was at least $5\dg$ from the (assumed known) desired source's direction.

The MUSIC algorithm is commonly used for narrowband DoA estimation, although extensions to broadband have been proposed \cite{alrmah_extension_2011}. For simplicity, we will apply it to each frequency bin, and employ two phasor-based averaging methods to obtain an angle estimate $\bar{\theta}$:

$\music$: In the first one, a simple average will be taken across frequency, such that
\begin{equation}
	\bar{\theta}_{\music} = \angle\sum_{k} \e^{\j\theta_{\music}[k]}.
\end{equation}

$\wmusic$: In the second one, a weighted average will be taken, weighing each bin by its MUSIC-estimated spectrum $p_{\wmusic}[k]$, such that
\begin{equation}
	\bar{\theta}_{\wmusic} = \angle\sum_{k} p_{\wmusic}[k] \e^{\j\theta_{w\music}[k]}.
\end{equation}

\subsection{Simulations for DoA estimation}
\label{subsec:sec6:sims_doa_estimation}

For the proposed DoA estimator, the initial guess is always $10\dg$--$15\dg$ away from the true source, emulating a spread-out multi-initialization process or a reasonable previous estimate.

\subsubsection{Evaluation criteria}

To compare the DoA estimation processes, we calculate the angular error between the estimate (through any of the shown methods) and the true interfering source's direction. This angular error is defined by
\begin{equation}
    \Delta \theta_{b;\condition} = \arccos\pts{\bvu[h]{\theta_b;\condition} \cdot \bvu[h]{\theta_b\opt;\condition}},
\end{equation}
wherein $\condition$ denotes any of the 1458 possible configurations of parameters as presented in \cref{tab:sec5:simulation_parameters}, and $\bvu[h]{\theta_b;\condition}$ is a unit vector with direction $\theta_b$, similarly for $\bvu[h]{\theta_b\opt;\condition}$. Given a parameter (among those in the first column of \cref{tab:sec5:simulation_parameters}) and a value within the possible ones for this parameter, the results' statistics will be presented as boxplots (showing median, interquartile range and $9$-th and $91$-th percentiles) where the value is fixed and we marginalize over all other parameters, with each possible parameter being represented in a different subfigure. This enables the partition of each parameter's effects, and to compare among the chosen values of each.

\begin{figure*}[t]%
    \def\fbox{}%
	\centering%
    \setlength{\parindent}{0pt}%
	\fbox{\begin{subfigure}{0.32\linewidth}%
		\centering%
		\begin{tikzpicture}%
			\begin{angleboxplot}[%
				xmin = 0,%
				xmax = 12,%
				xtick = {2, 6, 10},%
				xticklabels={$0\si{\milli\second}$, $500\si{\milli\second}$, $800\si{\milli\second}$},]%
                \addboxplot{0.36}{0.73}{1.56}{4.23}{5.08}[draw position=1][styleA]%
                \addboxplot{2.24}{3.78}{8.23}{12.34}{16.30}[draw position=2][styleC]%
                \addboxplot{0.75}{2.16}{4.83}{8.89}{13.74}[draw position=3][styleE]%
                %
                \addboxplot{0.57}{2.00}{4.40}{6.66}{7.95}[draw position=5][styleA]%
                \addboxplot{3.68}{6.26}{15.07}{28.51}{49.11}[draw position=6][styleC]%
                \addboxplot{2.15}{5.47}{14.21}{29.53}{53.72}[draw position=7][styleE]%
                %
                \addboxplot{0.55}{1.73}{4.14}{7.39}{8.71}[draw position=9][styleA]%
                \addboxplot{5.24}{8.70}{19.17}{38.59}{64.63}[draw position=10][styleC]%
                \addboxplot{2.89}{7.27}{18.53}{36.13}{62.15}[draw position=11][styleE]%
			\end{angleboxplot}%
		\end{tikzpicture}%
		\caption{Reverberation time $\Tsixty$ ($\si{\second}$)}%
		\label{fig:sec5:boxplot_prediction_error:subfig1}%
	\end{subfigure}}\hspace*{0.00\linewidth}%
	\fbox{\begin{subfigure}{0.32\linewidth}%
		\centering%
		\begin{tikzpicture}%
			\begin{angleboxplot}[%
				xtick = {2, 6, 10},%
				xmin = 0,%
				xmax = 12,%
				xticklabels={$50\si{\centi\meter}$, $150\si{\centi\meter}$, $300\si{\centi\meter}$},]%
                \addboxplot{0.44}{1.46}{4.15}{6.61}{8.14}[draw position=1][styleA]%
                \addboxplot{2.95}{5.57}{9.03}{16.19}{25.32}[draw position=2][styleC]%
                \addboxplot{1.70}{4.32}{11.15}{22.95}{52.25}[draw position=3][styleE]%
                %
                \addboxplot{0.43}{1.48}{4.37}{6.88}{8.30}[draw position=5][styleA]%
                \addboxplot{3.06}{6.77}{13.60}{28.42}{46.20}[draw position=6][styleC]%
                \addboxplot{1.59}{5.40}{12.82}{27.32}{55.81}[draw position=7][styleE]%
                %
                \addboxplot{0.35}{0.92}{2.55}{4.33}{6.29}[draw position=9][styleA]%
                \addboxplot{3.14}{7.05}{13.84}{33.23}{65.65}[draw position=10][styleC]%
                \addboxplot{1.03}{2.89}{7.15}{18.45}{39.20}[draw position=11][styleE]%
			\end{angleboxplot}%
		\end{tikzpicture}%
		\caption{Desired source's distance ($\si{\centi\meter}$)}%
	\end{subfigure}}\hspace*{0.00\linewidth}%
	\fbox{\begin{subfigure}{0.32\linewidth}%
		\centering%
		\begin{tikzpicture}%
			\begin{angleboxplot}[%
				xtick = {2, 6, 10},%
				xmin = 0,%
				xmax = 12,%
				xticklabels={$50\si{\centi\meter}$, $150\si{\centi\meter}$, $300\si{\centi\meter}$},]%
                \addboxplot{0.44}{1.80}{3.61}{5.23}{7.63}[draw position=1][styleA]%
                \addboxplot{3.34}{6.06}{11.67}{19.13}{32.79}[draw position=2][styleC]%
                \addboxplot{1.03}{3.71}{7.50}{13.89}{22.67}[draw position=3][styleE]%
                %
                \addboxplot{0.24}{0.93}{3.01}{5.65}{7.60}[draw position=5][styleA]%
                \addboxplot{3.64}{7.43}{11.35}{23.67}{48.91}[draw position=6][styleC]%
                \addboxplot{1.69}{4.31}{12.82}{29.44}{52.73}[draw position=7][styleE]%
                %
                \addboxplot{0.57}{1.12}{4.20}{6.87}{8.18}[draw position=9][styleA]%
                \addboxplot{2.73}{5.90}{11.77}{30.47}{60.54}[draw position=10][styleC]%
                \addboxplot{1.34}{3.86}{12.56}{35.65}{66.07}[draw position=11][styleE]%
			\end{angleboxplot}%
		\end{tikzpicture}%
		\caption{Interfering source's distance ($\si{\centi\meter}$)}%
        \label{fig:sec5:boxplot_prediction_error:subfig3}
	\end{subfigure}}\\[1em]%
	\fbox{\begin{subfigure}{0.32\linewidth}%
		\centering%
		\begin{tikzpicture}%
			\begin{angleboxplot}[%
				xtick = {2, 6, 10},%
				xmin = 0,%
				xmax = 12,%
				xticklabels={$-10\dB$, $0\dB$, $5\dB$},]%
                \addboxplot{0.19}{0.42}{1.47}{3.16}{4.81}[draw position=1][styleA]%
                \addboxplot{1.18}{2.94}{4.88}{7.08}{11.17}[draw position=2][styleC]%
                \addboxplot{0.58}{1.52}{3.70}{8.95}{19.31}[draw position=3][styleE]%
                %
                \addboxplot{0.73}{1.29}{3.42}{5.61}{6.75}[draw position=5][styleA]%
                \addboxplot{6.35}{9.00}{14.58}{27.51}{43.86}[draw position=6][styleC]%
                \addboxplot{2.72}{4.39}{11.59}{24.22}{48.00}[draw position=7][styleE]%
                %
                \addboxplot{2.64}{4.19}{5.92}{7.80}{8.63}[draw position=9][styleA]%
                \addboxplot{8.57}{13.91}{20.16}{39.91}{65.51}[draw position=10][styleC]%
                \addboxplot{5.45}{9.58}{19.46}{40.52}{69.11}[draw position=11][styleE]%
			\end{angleboxplot}%
		\end{tikzpicture}%
		\caption{Signal-to-Interference Ratio ($\dB$)}%
        \label{fig:sec5:boxplot_prediction_error:subfig4}%
	\end{subfigure}}\hspace*{0.00\linewidth}%
	\fbox{\begin{subfigure}{0.32\linewidth}%
		\centering%
    		\begin{tikzpicture}%
    			\begin{angleboxplot}[%
    				xtick = {2, 6, 10},%
    				xmin = 0,%
    				xmax = 12,%
    				xticklabels={$0\dB$, $5\dB$, $10\dB$},]%
                    \addboxplot{0.42}{1.34}{3.87}{6.00}{8.14}[draw position=1][styleA]%
                    \addboxplot{3.00}{6.32}{12.20}{23.57}{47.68}[draw position=2][styleC]%
                    \addboxplot{1.55}{3.61}{8.95}{20.99}{49.55}[draw position=3][styleE]%
                    %
                    \addboxplot{0.41}{1.25}{3.67}{5.89}{7.86}[draw position=5][styleA]%
                    \addboxplot{3.00}{6.25}{12.00}{22.67}{48.08}[draw position=6][styleC]%
                    \addboxplot{1.40}{4.26}{10.52}{22.42}{51.13}[draw position=7][styleE]%
                    %
                    \addboxplot{0.39}{1.20}{3.68}{5.92}{7.81}[draw position=9][styleA]%
                    \addboxplot{3.08}{6.15}{10.92}{22.46}{48.53}[draw position=10][styleC]%
                    \addboxplot{1.45}{4.31}{11.05}{24.22}{51.19}[draw position=11][styleE]%
    			\end{angleboxplot}%
    		\end{tikzpicture}%
    	\caption{Signal-to-Correlated Ratio ($\dB$)}%
	\end{subfigure}}\hspace*{0.00\linewidth}%
	\fbox{\begin{subfigure}{0.32\linewidth}%
		\centering%
    		\begin{tikzpicture}%
        		\begin{angleboxplot}[%
        			xtick = {2, 6, 10, 14, 18, 22},%
        			xmin = 0,%
        			xmax = 24,%
        			xticklabels={$10\dg$, $30\dg$, $50\dg$, $70\dg$, $90\dg$, $110\dg$},%
        			legend to name = {BoxPlotErrorDg},%
        			legend style={
        				legend columns=3,%
        				/tikz/every even column/.append style={column sep=1em}},%
        			legend cell align={left},%
        			]%
                    \addboxplot{0.36}{1.03}{2.82}{4.21}{6.21}[draw position=1][styleA]%
                    \addboxplot{0.38}{1.17}{2.66}{4.42}{14.88}[draw position=2][styleC]%
                    \addboxplot{0.35}{0.88}{2.45}{6.71}{19.87}[draw position=3][styleE]%
                    %
                    \addboxplot{0.41}{0.77}{1.98}{4.37}{6.66}[draw position=5][styleA]%
                    \addboxplot{1.11}{4.41}{8.09}{10.43}{18.73}[draw position=6][styleC]%
                    \addboxplot{0.30}{2.04}{5.37}{12.11}{18.86}[draw position=7][styleE]%
                    %
                    \addboxplot{0.14}{0.42}{2.52}{5.91}{7.55}[draw position=9][styleA]%
                    \addboxplot{2.35}{5.25}{11.66}{16.90}{31.55}[draw position=10][styleC]%
                    \addboxplot{0.83}{2.89}{8.29}{18.29}{27.79}[draw position=11][styleE]%
                    %
                    \addboxplot{0.38}{1.53}{3.97}{6.46}{7.92}[draw position=13][styleA]%
                    \addboxplot{3.20}{7.08}{14.16}{29.33}{44.04}[draw position=14][styleC]%
                    \addboxplot{1.58}{4.26}{11.47}{24.59}{45.14}[draw position=15][styleE]%
                    %
                    \addboxplot{1.10}{1.94}{4.21}{6.69}{8.29}[draw position=17][styleA]%
                    \addboxplot{3.67}{7.43}{15.81}{32.41}{55.03}[draw position=18][styleC]%
                    \addboxplot{2.72}{6.06}{14.40}{36.74}{56.70}[draw position=19][styleE]%
                    %
                    \addboxplot{0.81}{1.88}{4.37}{6.67}{8.49}[draw position=21][styleA]%
                    \addboxplot{3.92}{6.94}{17.01}{45.86}{70.91}[draw position=22][styleC]%
                    \addboxplot{2.68}{5.63}{16.07}{44.88}{78.54}[draw position=23][styleE]%
        			\addplot[styleA] coordinates{(-1, 1)};%
        			\addplot[styleC] coordinates{(-1, 1)};%
        			\addplot[styleE] coordinates{(-1, 1)};%
        			\addlegendentry{NCM};%
        			\addlegendentry{MSC};%
        			\addlegendentry{wMSC};%
        		\end{angleboxplot}%
        	\end{tikzpicture}%
		\caption{Interfering source's (true) DoA ($\dg$)}%
		\label{fig:sec5:boxplot_prediction_error:subfig6}%
	\end{subfigure}}%
	
	\vspace*{1em}%
	\ppref{BoxPlotErrorDg}%
    \caption{Angle prediction error statistics (boxplots), for parameters in \cref{tab:sec5:simulation_parameters}.}%
	\label{fig:sec5:boxplot_prediction_error}%
\end{figure*}
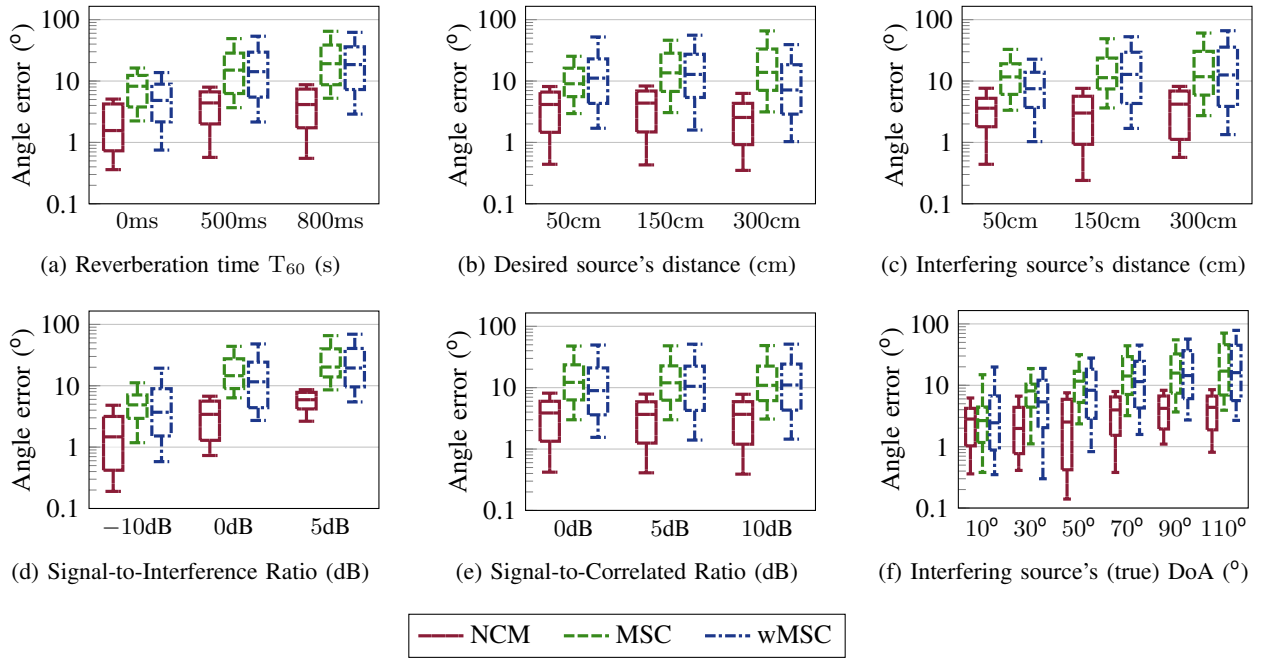%

\Cref{fig:sec5:boxplot_prediction_error} presents the DoA estimation error for the three presented approaches (proposed, standard average MUSIC, weighted average MUSIC), with each subfigure dissecting the performance along each parameter. The proposed method is denoted NCM (red), the standard MUSIC MSC (green, dashed), and the weighted MUSIC is denoted wMSC (in blue, dash-dotted). 
Given the extensive range of angle errors, spanning 4 orders of magnitude, the angular error is presented on a logarithmic scale. 

Across all plots in \cref{fig:sec5:boxplot_prediction_error}, the proposed technique consistently resulted in better DoA estimations, both in terms of median behavior as well as best- and worst-case outcomes (represented by the whiskers), mostly achieving errors below $5\dg$. The MUSIC-based DoA estimation errors are frequently $3$ times greater than the ones obtained through the proposed NCM-based approach.

Diving into the results for each fixed parameter, one sees that the DoA techniques' performance is most affected by $\Tsixty$, SIR, and true $\theta_b$ (respectively, \cref{fig:sec5:boxplot_prediction_error:subfig1,fig:sec5:boxplot_prediction_error:subfig4,fig:sec5:boxplot_prediction_error:subfig6}). While it is expected that both a higher SIR (meaning the interfering signal is less powerful) and a higher reverberation time would correlate to a higher DoA estimation error, since the directional signal to be estimated would be less present in the overall signal, we also see that the proposed technique is more robust to these parameters variations compared to the MUSIC-based methods.

Interestingly, the proposed technique's median performance is almost invariant regarding the interfering source's (true) DoA, with median performance fluctuating between $2$-$4\dg$ as seen in \cref{fig:sec5:boxplot_prediction_error:subfig6}. Meanwhile, the MUSIC-based performance deteriorates considerably as $\theta_b$ increases (from $2\dg$ to $17\dg$ error). Additionally, from \cref{fig:sec5:boxplot_prediction_error:subfig3} we see that the interfering source's distance to the array has little impact on the outcome, with little variation for all techniques across the presented values. This, coupled with the results in \cref{fig:sec5:boxplot_prediction_error:subfig6}, implies the proposed technique's robustness to interfering source's position, enabling both source tracking, as well as switching the interfering DoA between two (or more) interlocuting sources.

As would be expected, parameters unrelated to the interfering source, such as the desired source's distance and input SCR, have little to no impact on the estimators' performance.

\paragraph{Results under relaxed assumptions}%

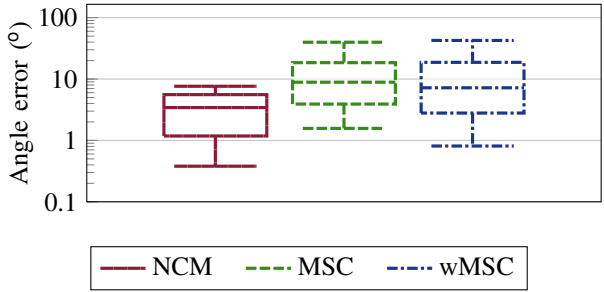
\begin{figure}[t]
	\centering
	\hspace*{-1em}
	\begin{tikzpicture}
		\begin{angleboxplot}[
			ylabel={Angle error ($\dg$)},
			xmin = 0,
			xmax = 4,
			xtick = \empty,]
			\addboxplot{0.38}{1.18}{3.43}{5.57}{7.63}[draw position=1][styleA]
            \addboxplot{1.57}{3.91}{8.88}{18.49}{39.78}[draw position=2][styleC]
            \addboxplot{0.81}{2.79}{7.20}{18.70}{42.57}[draw position=3][styleE]
		\end{angleboxplot}
	\end{tikzpicture}
	
	\vspace*{1em}
	\ppref{BoxPlotErrorDg}
	\caption{Angle prediction error statistics (boxplots), for situations with invalid assumptions.}
	\label{fig:sec5:boxplot_prediction_error_invalid_assumptions}
\end{figure}

Another class of scenarios that deserve some consideration are those where the assumptions from  \cref{sec:signal_model} aren't valid; namely, that the planar wave decomposition can't be applied, and that the coherent contaminating signals can't be modeled as an isotropic noise field. The violation of the aforementioned conditions applies when sources are close to the device, and/or a low reverberation time is coupled to a high SCR. Results obtained for this conditions are presented in \cref{fig:sec5:boxplot_prediction_error_invalid_assumptions}. Even in these adverse scenarios, the proposed technique still works and outperforms the other methods.

\paragraph{Discussion of general DoA estimation}

Overall, we see a considerable robustness difference between the proposed technique and the broadband-adapted MUSIC ones, with the NCM-based method consistently resulting in more accurate DoA estimations for a wide range of parameter variations and conditions.

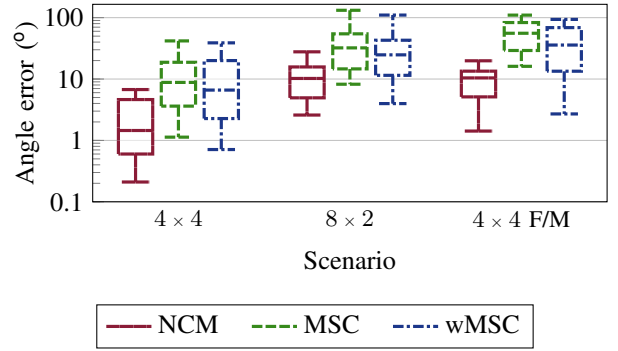
\begin{figure}[t]
	\centering
	\hspace*{-1em}
	\begin{tikzpicture}
		\begin{angleboxplot}[
			ylabel={Angle error ($\dg$)},
            xlabel={Scenario},
            xtick = {2, 6, 10},
            xmin = 0,
            xmax = 12,
            xticklabels={$\sz{4}{4}$, $\sz{8}{2}$, $\sz{4}{4}$ F/M},
    			]
            \addboxplot{0.21}{0.60}{1.45}{4.63}{6.74}[draw position=1][styleA]
            \addboxplot{1.13}{3.62}{8.81}{18.75}{41.81}[draw position=2][styleC]
            \addboxplot{0.71}{2.27}{6.61}{19.99}{38.87}[draw position=3][styleE]

            \addboxplot{2.59}{4.94}{10.22}{15.74}{27.76}[draw position=5][styleA]
            \addboxplot{8.23}{14.63}{32.19}{54.52}{131.89}[draw position=6][styleC]
            \addboxplot{3.98}{11.49}{24.78}{42.88}{110.21}[draw position=7][styleE]

            \addboxplot{1.42}{5.12}{10.43}{13.44}{19.83}[draw position=9][styleA]
            \addboxplot{16.11}{29.18}{55.71}{83.27}{110.00}[draw position=10][styleC]
            \addboxplot{2.70}{13.41}{35.80}{68.82}{93.66}[draw position=11][styleE]
		\end{angleboxplot}
	\end{tikzpicture}
	
	\vspace*{1em}
	\ppref{BoxPlotErrorDg}
	\caption{Global angle prediction error statistics, for each considered situation (base case, eccentric array, and transposed voice types), with simplified parameter set.}
	\label{fig:sec5:boxplot_prediction_error_all_cases}
\end{figure}

The proposed technique is also the top performer for an eccentric rectangular sensor array ($\sz{8}{2}$ rectangular array) and transposed directional sources' signals (desired female voice, and interfering male voice) conditions; both presented (along with the $\sz{4}{4}$ base-case) in \cref{fig:sec5:boxplot_prediction_error_all_cases}, for a sub-set of parameters from \cref{tab:sec5:simulation_parameters} ($d_x \in [150\cm,300\cm]$, $d_p \in [150\cm,300\cm]$, $\text{SIR} \in [-10\db, 0\db]$, $\text{SCR} \in [0\db,5\db]$; $\Tsixty$ and $\theta_b$ unchanged), totaling 288 combinations. It is also noticeable that a male interfering source (more low frequency energy, harder to identify direction \cite{chen_source_2002} and more susceptible to reverberation \cite{schroeder_natural_1962}), or a less symmetric array (has more accuracy in a preferred direction), both deteriorate the performance for all techniques, including the proposed one.

\subsection{Simulations for signal enhancement}

The performance of the selected beamformers (\cref{eq:sec4:beamformer_mvdr,eq:sec4:beamformer_lcmv,eq:sec4:beamformer_lcmp}) will be assessed for the scenarios given by the parameters in \cref{tab:sec5:simulation_parameters}, with a $\sz{4}{4}$ array, a male-voice desired signal, and a female-voice interfering signal; conditions identical to those originally analyzed in \cref{subsec:sec6:sims_doa_estimation}.

\subsubsection{Evaluation criteria}
Different metrics are used to evaluate the performance of the selected beamformers (\cref{eq:sec4:beamformer_mvdr,eq:sec4:beamformer_lcmv,eq:sec4:beamformer_lcmp}). For all metrics, we consider the desired signal as only the desired source signal's direct-path portion. All considered metrics are computed for the signals in the discrete time domain. The subscript $(\cdot)_{\filt}$ denotes the filtered form of the respective signal; that is, $x_{\filt}[l,k] = \he{\bvh}[l,k] \bvx[l,k]$, and the same applies to all other filtered signals. For clarity, all metrics and their statistics will be presented on a logarithmic scale ($\dB$). 

To evaluate signal enhancement, the gains in SIR and SNR\footnote{Here, SNR will represent the ratio w.r.t. the global noise $\eta$.} (denoted gSIR and gSNR), and the interfering signal reduction factor (ISRF) will be used, these being respectively defined as
\begin{subgather}
	\gsnr = \frac{\var{x_{\filt}}}{\var{\eta_{\filt}}} \frac{\var{\eta_1}}{\var{x_1}},\\
	\gsir = \frac{\var{x_{\filt}}}{\var{n_{\filt}}} \frac{\var{n_1}}{\var{x_1}},\\
	\isrf = \frac{\var{p_1}}{\var{p_{\filt}}}.
\end{subgather}

While $\gsnr$ and $\gsir$ measure the improvement in signal-to-contamination ratio, $\isrf$ reflects the interfering source's signal direct-path rejection. Note that $\gsir$ represents the power-ratio between the observed signal's direct-path component, and the interfering signal $n_1(t)$, encompassing all of the latter's planar waves.

To characterize the desired signal's preservation, the desired signal reduction factor (DSRF)
\begin{equation}
	\dsrf = \frac{\var{x_1}}{\var{x_{\filt}}} 
\end{equation}
will be used. Although desired signal distortion index (DSDI) is also commonly used for assessing desired signal preservation, DSRF was chosen for its $\dB$-friendly quality. We highlight that both DSRF and ISRF are measured w.r.t. the direct-path (or main planar wave) portion of the signals, disregarding reverberations and late-arrivals.

Lastly, theoretical performance assessment is done through the broadband directivity factor (DF) 
and white-noise gain (WNG), defined as
\begin{subgather}
	\df = \frac{K}{\sum\limits_{k=0}^{K} \he{\bvh}[k] \bvGa[k] \bvh[k]}, \\
	\wng = \frac{K}{\sum\limits_{k=0}^{K}\he{\bvh}[k] \bvh[k]},
\end{subgather}
being $\bvGa[k]$ the isotropic noise field pseudo-covariance matrix \cite{epain_spherical_2016,habets_generating_2007}. These metrics assess the beamformer's rejection to an ideal isotropic noise field and a purely Gaussian noise, respectively.

The following three beamformers will be compared:
\begin{itemize}
	\item LCMV: with proposed DoA and NCM estimation (red, solid line);
	\item MVDR: only the modeled NCM (yellow, densely dotted);
	\item LCMP: using the MUSIC algorithm (green, dashed);
\end{itemize}

Since both standard and weighted average MUSIC estimators had almost identical DoA error performance (\cref{fig:sec5:boxplot_prediction_error,fig:sec5:boxplot_prediction_error_invalid_assumptions,fig:sec5:boxplot_prediction_error_all_cases}), only the standard average MUSIC algorithm will be used.

\subsubsection{Results and discussion}

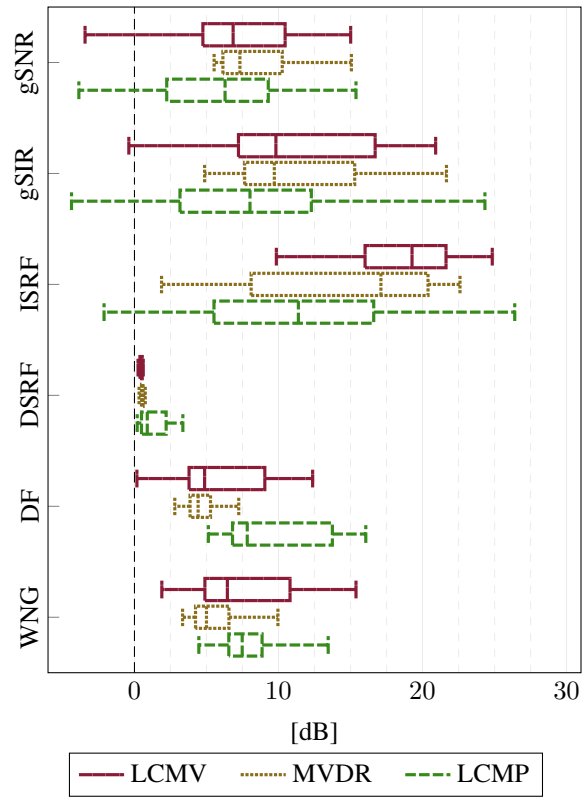
\begin{figure}[t]
	\centering
	\hspace*{-1em}
	\begin{tikzpicture}
		\begin{boxplot}[
			height=1\linewidth,
			width=0.8\linewidth,
			scale only axis,
			ymin = 0, ymax = 24,
			ytick = {2, 6, 10, 14, 18, 22},
			yticklabels = {WNG,DF,DSRF,ISRF,gSIR,gSNR},
			yticklabel style={
				rotate=90,
				anchor=south,
				align=center,
				text width=5em
				},
			%
			xlabel={[$\dB$]},
			xmin=-6, xmax=31,
			xtick={0, 10, 20, 30},
			boxplot/draw direction = x,
			legend to name = {BoxPlotGainSNR},
			legend style={
				legend columns=3,
				/tikz/every even column/.append style={column sep=1em}
			},
			legend cell align={left},
			xmajorgrids, xminorgrids,
			grid style={line width=.1pt, draw=gray!15},
			minor x tick num=3,minor grid style={line width=.1pt, draw=gray!15, dashed},
		]
			\addplot[forget plot, thin, dash pattern = {on 4pt off 2pt}] coordinates{(0,0) (0,17.5-0.1)};
			\addplot[forget plot, thin, dash pattern = {on 4pt off 2pt}] coordinates{(0,35) (0,17.5+0.1)};
			%
			%
			
			\addboxplot{-3.43}{4.75}{6.85}{10.46}{15.01}[draw position=23][styleA]          
            \addboxplot{5.53}{6.14}{7.32}{10.26}{15.06}[draw position=22][styleB]           
            \addboxplot{-3.86}{2.24}{6.29}{9.29}{15.38}[draw position=21][styleC]           
            
            \addboxplot{-0.39}{7.22}{9.82}{16.71}{20.91}[draw position=19][styleA]          
            \addboxplot{4.87}{7.65}{9.71}{15.29}{21.66}[draw position=18][styleB]           
            \addboxplot{-4.37}{3.16}{8.02}{12.28}{24.33}[draw position=17][styleC]          
            
            \addboxplot{9.85}{16.01}{19.27}{21.63}{24.84}[draw position=15][styleA]         
            \addboxplot{1.88}{8.10}{17.11}{20.38}{22.60}[draw position=14][styleB]          
            \addboxplot{-2.11}{5.52}{11.38}{16.61}{26.40}[draw position=13][styleC]         

            \addboxplot{0.25}{0.38}{0.44}{0.54}{0.61}[draw position=11][styleA]             
            \addboxplot{0.32}{0.49}{0.55}{0.60}{0.75}[draw position=10][styleB]             
            \addboxplot{0.19}{0.49}{0.89}{2.20}{3.36}[draw position=9][styleC]              
            
            \addboxplot{0.17}{3.79}{4.87}{9.05}{12.37}[draw position=7][styleA]             
            \addboxplot{2.80}{3.86}{4.42}{5.28}{7.24}[draw position=6][styleB]              
            \addboxplot{5.13}{6.81}{7.83}{13.74}{16.06}[draw position=5][styleC]            
            
            \addboxplot{1.90}{4.89}{6.45}{10.80}{15.38}[draw position=3][styleA]            
            \addboxplot{3.34}{4.24}{5.00}{6.56}{9.96}[draw position=2][styleB]              
            \addboxplot{4.47}{6.55}{7.48}{8.86}{13.45}[draw position=1][styleC]             
			
			\addplot[styleA] coordinates{(0, -1)};
			\addplot[styleB] coordinates{(0, -1)};
			\addplot[styleC] coordinates{(0, -1)};
			\addlegendentry{LCMV};
			\addlegendentry{MVDR};
			\addlegendentry{LCMP};
		\end{boxplot}
	\end{tikzpicture}
	\vspace*{1em}
	\ppref{BoxPlotGainSNR}
	\caption{Beamformers' performance, measured with different evaluation criteria (gSNR, gSIR, ISRF, DSRF, DF and WNG).}
	\label{fig:sec5:boxplot_bf_metrics}
\end{figure}

We present the statistics for each evaluation criteria and beamformer in \cref{fig:sec5:boxplot_bf_metrics}, for the same simulated conditions given by the parameters in \cref{tab:sec5:simulation_parameters}. In the discussion, lower- and upper-bounds refer to the $9$-th and $91$-th whiskers; and lower- and upper-quartiles to the $25\%$ and $75\%$ quartiles.

\paragraph{Noise signal rejection}

In terms of global noise rejection measured by $\gsnr$, we see that all techniques fared almost identically in terms of median outcome, with some differences in expected (quartiles) and extreme-case (whiskers) scenarios. A similar result is found in terms of $\gsir$.

The first main difference can be seen in terms of interfering source's direct-path rejection (measured by $\isrf$), in which the proposed technique consistently outperformed the MUSIC algorithm, and marginally the MVDR beamformer, in all statistical measures represented in the boxplots. This is linked to the directional null present in the LCMV formulation, which isn't used in the MVDR. Interestingly, the MVDR had better direct-path interfering signal rejection than the LCMP, which also contains a null steering step in its beamformer.

\paragraph{Desired signal preservation}

The DSRF results indicate that the NCM-based beamformers exhibit lower desired signal distortion compared to their MUSIC-based counterpart, specially in terms of upper-quartile and upper-bound performance; exposing the proposed approach's consistency. This improvement is attributed to the proposed NCM design, which excludes the desired signal from the minimized covariance matrix, minimizing signal distortion. This desired signal distortion is also marginally linked to the slightly worse performance of the MUSIC beamformer in terms of gSNR and gSIR, as these gains can be defined based on signal reductions factors \cite{benesty_fundamentals_2017}.

\paragraph{Theoretical noise field metrics}

Regarding the directivity factor we see that the standard average MUSIC-based LCMP had a convincingly superior performance compared to the NCM-based beamformers; and a slightly better result in terms of WNG, for lower-bound to median metrics. This may imply that the MUSIC-based DoA estimation is better for fixed beamforming applications, although this assertion would need more exploration.

\paragraph{General discussion on signal enhancement}

From \cref{fig:sec5:boxplot_bf_metrics} it is evident that an LCMV beamformer employing the proposed framework achieved an improved direct-path interfering signal rejection (via $\isrf$), while maintaining comparable results in direct-path signal rejection (through $\gsir$) and total interference reduction (with $\gsnr$). The improved direct-path rejection in the LCMV is primarily linked to the more precise DoA estimates provided by our proposed scheme (compared to the MUSIC algorithm, \cref{subsec:sec6:sims_doa_estimation}), and the absence of directional constraints in the MVDR. Conversely, the noise reduction validates the effectiveness of the proposed broadband NCM scheme's use for beamforming.

The NCM-based beamformers (both LCMV and MVDR) also exhibited a lower, and less variant, level of desired signal direct-path distortion (quantified by the $\dsrf$), when compared to the LCMP beamformer. This directly correlated to their common reliance on the NCM model, which inherently excludes the desired signal from the covariance matrix used in the filter construction, reducing the distortion in the desired signal.

While the beamformers based on our developed framework demonstrated superior performance in the practical signal enhancement metrics, the LCMP filter showed best results in the theoretical noise field scenarios, particularly in the isotropic noise field case (evaluated by the $\df$ metric).

\section{Conclusion}
\label{sec:conclusion}

The proposed work explored a joint estimation framework for the Noise Covariance Matrix (NCM) and Direction-of-Arrival (DoA), based on planar-wave decomposition and a broadband cost function. This method built upon a previously proposed NCM model by incorporating a novel broadband DoA estimator and improving the original optimization scheme. This revision enables a more direct and efficient estimation pipeline, yielding a quasi-linear solution (rather than an exhaustive search) for the NCM estimation part of the problem. Key advantages of our approach include its iterative DoA search, broadband formulation, and independence from a voice activity detector.

Through simulations emulating real-life environments, we demonstrated that the method achieves accurate DoA estimates, particularly in wider angle scenarios where conventional approaches often degrade. We benchmarked the proposed technique against state-of-the-art approaches described in the literature, varying sensor array configurations and signal types, with our novel approach consistently outperforming existing methods in both scenarios, exhibiting superior robustness.

Furthermore, we applied the estimated parameters to beamforming, where our proposed model (using the Linearly Constrained Minimum Variance formulation) presented superior direct interference suppression and desired signal preservation, when compared to beamformers with literature-based DoA estimation, while achieving similar general noise minimization.


%
\printbibliography[]
\appendix
\crefalias{subsection}{appendix}
\renewcommand{\theequation}{\thesubsection.\arabic{equation}}
\setcounter{equation}{0}

\subsection{Proofs and Theorems}
\label{app:appB:proofs}
Let the function $\cost*{\bvsi} \equiv \cost*{\Corr{\bvy[t]}(\bvsi{k},\Theta)}[k]$ from \cref{eq:sec3:cost_function} be the function to be minimized w.r.t. $\bvsi$, under $\bvsi \geq \bv0$.

Given that $\cost*{\bvsi}$ is quadratic on any element of $\bvsi$ (obtainable from \cref{eq:sec3:cost_function_longform}), it is guaranteed to have a global extrema, denoted $\bvsi\opt$, which may not respect the constraints. Since $\cost*{\bvsi}$ is the sum of squared real numbers, its extrema is a minimum. Given the triviality of scenarios where $\bvsi\opt \geq \bv0$, we assume that $\bvsi\opt \not>\bv0$, having at least one negative entry. We let $\bvsi\opt*$ be the minimum of $\cost*{\bvsi}$ which fulfills $\bvsi\opt* \geq 0$; and define $\bvsi(t) = \bvsi\opt + (\bvsi\opt* - \bvsi\opt)t$ as a solution parameterized by $t$; trivially, $\bvsi(0) = \bvsi\opt$, and $\bvsi(1) = \bvsi\opt*$. We also define $J(t) \equiv \cost*{\bvsi\opt + (\bvsi\opt* - \bvsi\opt)t}$.

\begin{theorem}{If the global minimum isn't in the feasible region, the constrained minimum lies in its boundary.}[\label{thm:constrained_min_boundary}]
	
	We assume that $\bvsi\opt* > \bv0$, strictly greater; that is, the constrained minimum isn't on the boundary where some entries are $0$. Through the intermediate value theorem, there exists some $t' \in [0,1]$ for which $t \geq t'$ implies that $\bvsi(t) \geq \bv{0}$ (greater or equal).
	
	Using $J(t)$ as defined previously using the definition from \cref{eq:sec3:cost_function_longform}, replacing $\bvsi = \bvsi\opt + (\bvsi\opt* - \bvsi\opt)t$, and expanding, leads to
	\begin{equations}
		J(t)
		& = \sum_{\bvj} \abs{\bts{\sum_{z\in[x, u, \gamma, v]} R_{\bvz;\bvj} \sigma_z(t)} - R_{\bvy,\epsilon;\bvj}}^2 \\
		& = \sum_{\bvj} \pts{\Re{a_{\bvj}}t - \Re{b_{\bvj} }}^2 + \pts{\Im{a_{\bvj}}t - \Im{b_{\bvj} }}^2 \\
		& = At^2 + Bt + C,
	\end{equations}
	with
	\begin{subgather}
		A = \sum_{\bvj} \abs{a_{\bvj}}^2, \\
		B = -2\sum_{\bvj} \Re{a_{\bvj}} \Re{b_{\bvj}} + \Im{a_{\bvj}} \Im{b_{\bvj}} ,\\
		C = \sum_{\bvj} \abs{b_{\bvj}}^2 , \\
		a_{\bvj} = \sum_{z\in[x,u,\gamma,v]} R_{\bvz;\bvj}(\sigma_{z}\opt* - \sigma_{z}\opt) ,\\
		b_{\bvj} = R_{\bvy,\epsilon;\bvj} + \sum_{z\in[x,u,\gamma,v]} R_{\bvz;\bvj}(\sigma_{z}\opt*).
	\end{subgather}
	
	Note that $A > 0$, indicating that $J(t)$ is a positively-curved quadratic function in $t$. By definition, $J(0)$ minimizes the function, so $J(t) \geq J(0)~\forall~t$, and in particular $\forall~t\in[0,1]$. Since $J'(t)_{t=0} = 0$, we also have that $B = 0$. So, for any $0 < t' \leq t$
	\begin{equations}
		J(t')
		& = A(t')^2 + C \\
		& < At^2 + C = J(t),
	\end{equations}
	Since $t'$ is the smallest value for which $J(t') \geq 0$, and $t' < 1$, then $J(t') < J(1) = \bvsi\opt*$, contradicting the premise that $\bvsi\opt*$ is the optimal solution within our constraint space. Therefore, $\bvsi\opt* \not>\bv0$, and the optimal solution has (at least) one zero entry.
\end{theorem}

\begin{theorem}{The constrained entries of $\bvsi\opt*$ are at most the negative entries of $\bvsi\opt$.}
	\label{thm:no-constraints_global_min_positive_values}
	Since $\bvsi\opt* \geq 0$, and $\bvsi(t)$ is linear in $t$, the entries of $\bvsi\opt$ which are greater than $0$ will be positive $\forall t$. 
	This implies that any positive entry of $\bvsi(0) = \bvsi\opt$ will need not be constrained in the search for $\bvsi(1) = \bvsi\opt*$, being strictly positive along the curve defined by $\bvsi(t)$, $t\in[0,1]$. This, along with \cref{thm:constrained_min_boundary}, implies that only the negative entries of $\bvsi\opt$ may have active constraints.
	
	We remark that not all negative entries $\bvsi\opt$ must be set to $0$. Each constraint added changes the matrix $\bvA[k]$ from \cref{subeq:sec3:definition_bvA,eq:sec3:solution_unconstrained-minim_bvsi}, changing the manifold over which the minimization is done, resulting in a different minimum where the other negative entries of $\bvsi\opt$ may be naturally positive.
\end{theorem}
\subsection{List of symbols}
\label{app:appA:list_of_symbols}
Due to the work's heavy mathematical nature, we present a list of the most relevant symbols and variables used.

We again note that non-bold variables are scalars, bold lowercase variables are vectors, and bold uppercase variables are matrices. The placeholder variables $z$ and $\bvz$ can represent either of the previously presented signals.

\begin{tabular}{l l}
	$\bvx[l,k]$ & Desired signal vector \\
	$\bvp[l,k]$ & Interfering directional signal vector \\
	$\bvga[l,k]$ & Isotropic noise vector \\
	$\bvv[l,k]$ & White noise vector \\
	$\bvy[l,k]$ & Observed signal vector \\
	$\bveta[k]$ & Global noise vector \\
	$z_m,w_m$ & Token signals $z$ and $w$ at $m$-th sensor \\
	$\bvz,\bvw$ & Token signal vectors $z$ and $w$ \\
	$\Corr{\bvz}[k]$ & Covariance matrix for $\bvz$ \\
	$\var{z}[k]$ & Variance of the signal $z$ \\
	$\Corr{\bvy[t]}[k]$ & Observed signal estimate CM\\
	$\theta_b$ & Interfering source's azimuth \\
	$\phi_b$ & Interfering source's elevation \\
	$\Theta_b = (\theta_b,\phi_b)$ & Interfering source's DoA \\
	$\bvsi[k]$ & Variance vector \\
	$\bvSi$ & Cross-frequency variance matrix \\
	$\cost{\bvSi,\Theta_b}$ & Global cost function \\
    $\lag(\bvsi,\Theta_b,\bvze,\bvmu)$ & Lagrange multiplier function \\
	$\bvze[k]$ & Lagrange multiplier vector\\
	$\bvh[k]$ & Beamformer vector \\
	$\epsilon$ & Regularization factor \\
	$C[k]$ & Constraint matrix for LCMV filter
    \end{tabular}
%
%

\end{document}